\documentclass{aa}

\usepackage{graphicx,amsmath,multirow,amssymb}
\usepackage{natbib}
\usepackage{txfonts}

\newcommand{\msun}{\ensuremath{\, {M}_\odot}}
\newcommand{\Msun}{\ensuremath{\, {M}_\odot}}

\newcommand{\lsim}{\ \raise -2.truept\hbox{\rlap{\hbox{$\sim$}}\raise 5.truept\hbox{$<$}\ }}
\newcommand{\gsim}{\ \raise -2.truept\hbox{\rlap{\hbox{$\sim$}}\raise 5.truept\hbox{$>$}\ }}

\begin{document}

\title{Multipopulation aftereffects on the color-magnitude diagram
  and Cepheid variables of young stellar systems}
 \titlerunning{Multipopulation aftereffects in young stellar systems}
%   \subtitle{Multi-populations in young stellar systems}

   \author{R. Carini
          \inst{1}
          \and
          E. Brocato\inst{1}
                    \and
          M. Marconi\inst{2}
          \and
          G. Raimondo\inst{3}
                   }

                   \institute{INAF-Osservatorio Astronomico di Roma, via Frascati 33, 00040 Monte Porzio Catone, Italy \\
                     \email{roberta.carini@oa-roma.inaf.it,
                       enzo.brocato@oa-roma.inaf.it} \and
                     INAF - Osservatorio Astronomico di Capodimonte, Salita Moiariello 16, I-80131 Napoli, Italy\\
                     \email{marconi@na.astro.it} \and
                     INAF - Osservatorio Astronomico di Teramo, Mentore Maggini s.n.c., 64100 Teramo, Italy \\
                     \email{raimondo@oa-teramo.inaf.it}
%             \thanks{}
             }
             
    \abstract
   {The evidence of a multipopulation scenario in Galactic globular clusters  raises several questions about the
     formation and  evolution of the two (or more) generations of
     stars. These populations show differences in  their age and  chemical composition. These differences are
     found in old- and intermediate- age stellar clusters in the Local
     Group. The observations of young stellar systems are
     expected to present footprints of multiple
     stellar populations.  }
   {This theoretical work intends to be a specific step in exploring
     the space of the observational indicators of multipopulations,
     without covering all the combinations of parameters that may
     contribute to the formation of multiple generations of stars in a
     cluster or in  galaxy. The goal is to shed light on  the possible
     observational features expected by core He-burning stars that
     belong to two stellar populations with different original He
     content and ages. }
   {The tool adopted was the stellar population synthesis. We used new
     stellar and pulsation models  to construct a
     homogeneous and consistent framework. Synthetic color-magnitude
     diagrams (CMDs) of young- and intermediate-age stellar systems
     (from 20 Myr up to 1 Gyr) were computed in several photometric
     bands to derive possible indicators of double populations both in
     the observed CMDs and in the pulsation properties of the
     Cepheids.}
   {We predict that the morphology of the red/blue clump in VIK bands
     can be used to  photometrically  indicate the 
     two stellar populations in a rich assembly of stars if there is a
     significant difference in their original He content.  Moreover,
     the period distribution of the Cepheids appears to be  widely affected by
     the coeval multiple generations of stars within 
     stellar systems. We show that the Wesenheit relations may
     be affected by the helium content of the Cepheids.}
   {}
   % {This theoretical work intends to be a specific step in exploring
   %   the space of the observational indicators of multi-populations,
   %   without covering all the combination of parameters that may
   %   contribute in the formation of multiple generation stars in a
   %   cluster or in a galaxy. }


   \maketitle
           \section{Introduction}

A large number of young- and intermediate-age massive clusters (YIMCs) are
known in the local Universe as far away  as among nearby star-burst galaxies.
These systems are one of the preferred environments for star
formation to take place \citep{lel}. For this reason, they may be responsible for
a signi\-ficant fraction of the current star formation in the local
Universe and be the intriguing progenitors of the present-day population
of old globular clusters (GCs).  Therefore, to study YIMCs in terms of
their stellar content is crucial for understanding both current and
primordial star formation processes in galaxies.

In the past decade, our knowledge on star formation in GCs has been
revolutionized. High-resolution spectroscopy and precise photometry of
individual stars in GGCs provided the
evidence that multiple stellar generations are present
in these stellar systems \citep[e.g.][]{grat2012,piotto2010}.  In the
Magellanic Clouds (MCs), stellar clusters show a wide spread in age
(e.g. \citealt{brob} and reference therein), and signs of two or more
stellar populations have been found in young- and intermediate-age objects
\citep[]{mac,glatt,milone09, goud, keller}, suggesting that a star cluster can host
(at least) two stellar populations with different ages ($\Delta t < 1$ Gyr) and chemical compositions.

The current view on the matter suggests that, generally, the second generation (SG)
of stars differs from the first generation (FG) mainly in the He
abundance, and this difference can be as high as $\Delta Y \simeq 0.1$
\citep[e.g.][]{hb,piotto}. Two main models have been proposed to
explain the presence of multiple stellar generations in star clusters and their chemical abundances:
the so-called super-asymptotic giant branch (S-AGB) and AGB stars
scenario \citep{ventura2001}, and the fast-rotating massive stars
scenario \citep[FRMS,][]{decressin}.\\
According to the first scenario, during the AGB evolution stars with
mass $M \gsim 4 \msun$ experience the so-called hot bottom burning
(HBB) at the bottom of their convective envelope \citep{shon}. The H-rich matter is processed through the hot CNO cycle, 
where the Ne-Na
and Mg-Al chains are active. The processed material is ejected into
the intracluster medium and retained in the potential well of the
cluster, due to the small velocities of the AGB stellar winds. The
enriched matter mixes with the pristine gas and forms the second-generation stars
\citep{dercole2008}. The evolutionary timescales of this model range
from $\delta t \approx 40$ Myr to $\delta t \approx 100$ Myr. The
former timescale is driven by the evolution time of the most massive
S-AGB stars ($M \sim 8 \msun$\footnote{This mass value represents the limit between S-AGB and stars that explode as Type II
  supernovae (SNII). The value decreases at low metallicity and for
  high efficiency of the overshooting mechanism.}), while the latter
is related to the lifetime of stars with $M \sim 4 $\msun,
which also corresponds  to the minimum time needed for Type Ia supernovae (SNe) to
occur. These events add  positive energy to the system, stopping the star formation process.

The FRMS scenario suggests that during the main sequence (MS), massive
stars ($M = 20-60 $\msun) can eject CNO-processed
material at low velocity, transported from the inner regions to the surface by
rotationally induced mixing: the required rotation rates are close to
the break-up velocity. If an accretion disk is formed, the gas ejection
velocity is extremely low, thus the processed material can be
retained in the potential well of the cluster and  can be mixed
with the pristine gas, forming the second-generation stars. In this case, the
evolution time ranges from 5 Myr to $\sim$ 20 Myr and ends before Type II SNe
explosions occur.

These two models are not yet fully accepted, since they lack a
self-consistent explanation of the picture, and weaknesses are still
affecting the two proposed scenarios. For example, uncertainties
affect the number ratio between the first- and second-generation stars. 
Recently, \cite{bastian2} proposed a new picture where 
low-mass pre-main sequence  stars accrete enriched material released from interacting massive 
binary and rapidly rotating stars onto their circumstellar disks, and ultimately onto the young stars. 

In this context, a relevant
contribution comes from the analy\-sis of young stellar systems
(e.g. $t < 1$ Gyr) where possible second-generation stars are newly formed.
Searches of multiple stellar populations have been mainly conducted in old
GCs of the Galaxy. In old systems, multiple populations are typically
identified from the photometry of main sequence, red giant branch (RGB) and horizontal branch (HB) stars
\citep[e.g.][]{hb,bedin}. However, high-resolution photome\-tric data are required 
 to distinguish multiple turn-off (TO) and to evaluate the age difference
between the first- and  second-generation stars if younger than about 1 Gyr.
In contrast, in young- and intermediate-age
GCs even a small age difference  can produce multiple TO or other
easily  detected photometric signatures. Multipopulations have been
discovered by fitting the location of the MS in the observed CMD
\citep{milone}. Moreover, since the clump of He-burning stars is more
luminous than the MSTO, it potentially represents a more effective way
to recognize the presence of multipopulations in young massive
clusters.  For these reasons, YIMCs are systems that can be very powerful
to help in understanding the formation processes of the second-generation
stars and, more importantly, they provide a key opportunity to
distinguish between the nature of the polluters responsible for the new
 chemical abundances of second-generation stars, thus
giving solid constraints in favor or against the proposed scenarios.
On the base of visual inspection of spectra and CMDs of 130 YIMCs, \cite{bastian} 
suggested that models with continuous star formation are ruled out, while models requiring the nearly instantaneous formation of a 
secon\-dary population are not formally discounted. In their analysis the disk accretion model 
\citep{bastian2} seems to be preferred.

In this paper, we tackle this open question by studying the 
expected photometric properties of resolved
stars in young- and intermediate-age stellar clusters under the
hypothesis of single-burst and multipopulation scenarios. Our goal is
to investigate the features of
core He-burning stars in the CMD  and
the observational properties of  the Cepheids expected in two simple cases: i)
single-burst stellar populations where all stars have the same age and a
chemical composition  typical of the Large Magellanic Cloud (LMC), i.e. Y=0.25 and Z=0.008; ii) stellar
systems where first-generation stars (Y=0.25, Z=0.008 and age $ t = t_{FG}$)
are mixed with a second generation of stars born after a time $\delta t =
t_{FG}-t_{SG}$ (ranging from 20 Myr up to 100 Myr) that have a different
He abundance (Y=0.35). The age of first-generation stars ranges
from a few Myr up to $\sim$ 1 Gyr. The idea is to explore and point out
differences in the photometric  properties of stars expected in the two cases, by taking advantage of stellar population
synthesis methods. For this purpose, we derived synthetic CMDs  in the
classical Johnson-Cousins photometric bands (UBVRIJHK).

Since the Cepheid variables are an important stellar component of young
clusters, we  predict their properties using po\-pulation synthesis models coupled with high precision pulsation models. We  evaluate the impact of
multipopulations on the
pulsation properties such as the period distribution and the Wesenheit index.
 We aim to point out possible bias in the estimate of astronomical distances
based on Cepheid relationships.

The paper is arranged as follow: Section 2 describes the new
stellar evolution models, the new
stellar pulsation models and  briefly introduces of the adopted stellar
population synthesis code. The results of the synthetic CMDs are
discussed in Sect. 3, while the prediction on the properties for the Cepheids
and the impact on the Wesenheit relations are shown in Sect. 4. A
brief discussion  closes the paper.
\section{Physical and numerical inputs}
%                                     Two column figure (place early!)
%______________________________________________ Gamma_1 (lg rho, lg e)
\label{modello}
\begin{figure}[t]
\center
%\includegraphics[width=1\columnwidth,natwidth=600,natheight=650]{tracce.pdf}
%\vspace{-3cm}
\includegraphics[width=1\columnwidth,natwidth=750,natheight=0]{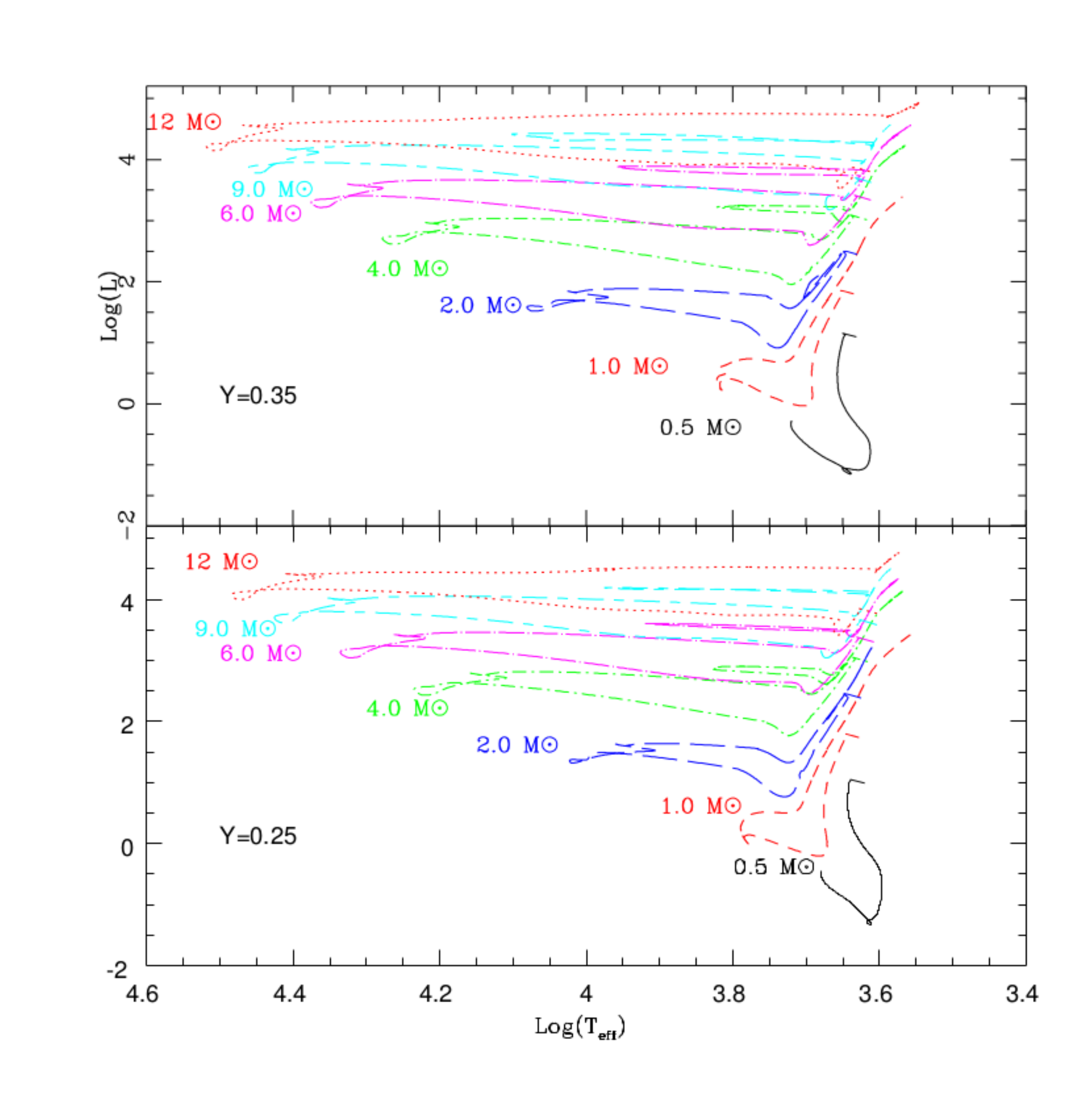}
%\includegraphics[width=0.45\textwidth]{tracce.pdf}
%\vspace{-0.7cm}
\caption{Selected sample of evolutionary tracks calculated with the ATON code. The initial
  chemistry is:   Z=0.008,
  $\alpha$-enhancement [$\alpha$/Fe]=0.2, and Y=0.25 (bottom panel)
  and Y=0.35 (top panel).}
\label{traccerob}
\end{figure}

\subsection{Evolution code}

The stellar evolution code used here is ATON. This  can follow all the
evolutionary phases from the pre-MS to D-ignition at the beginning of
the MS, up to carbon ignition \citep{aton}. The ignition of the
reaction $^{16}O$ + $^{16}O$ and the following pre-supernova phases
are not accounted for. In the following we
briefly recall the main features of the code.

ATON describes spherically symmetric
structures in hydrostatic equilibrium. The internal structure is
integrated via the Newton-Raphson relaxation method from the center
to the base of the optical atmosphere ($\tau$=2/3).  The independent
variable is the mass, while the dependent ones are temperature,
pressure, radius, and luminosity.  The zoning is reassessed after each
phy\-sical time step, with particular care to the central and surface
regions in the vicinities of convective boundaries and H or
He-burning shells and close to the superadiabacity zones.

Once the physical structure of the model is determined via the
relaxation method, a time step is applied to achieve the che\-mical
evolution.  To this end, each physical time step is divided in to ten
chemical time steps. Mechanisms leading to a variation of
the chemical composition, such as nuclear burning, gravitational settling,
and convective mixing, are properly taken into account.  All the
evolution models presented here  were calculated by adopting the
full spectrum of turbulence (FST) treatment  with the appropriate convective flux distribution
\citep{canuto}. Mixing of the chemicals within convective zones is
treated as a diffusive process: for an individual element $X_i$ the
code solves the equation by \citet{cloutman}:
\begin{equation}
 \frac{dX_i}{dt}= \frac{\partial X_i}{\partial t} + \frac{\partial}{\partial m_r}\left[(4\pi r^2\rho^2)^2 D\frac{\partial X_i}{\partial m_r}\right],
\end{equation}
where $D$ is the diffusion coefficient ($D=16\pi r^4 \rho^2
\tau^{-1}$), $\rho$ is the density, and $\tau$ is the turbulent diffusion
timescale. The mass fraction of the element is defined as $X_i$ =
$X_i/\rho N_A$, where $N_A$ is the Avogadro's number. A network of 64 reactions is adopted. The cross sections are taken  from the NACRE
compilation \citep{angulo}, with the following exceptions:

\begin{itemize}
 \item $^{14}N(p,\gamma)^{15}O$ \citep{formicola}
 \item $^{22}Ne(p,\gamma)^{23}Na$ \citep{hale02} (upper limit)
 \item $^{23}Na(p,\gamma)^{24}Mg$ \citep{hale02}
 \item $^{23}Na(p,\alpha)^{22}Ne$ \citep{hale02} (lower limit)
 \item $^{4}He(2\alpha,\gamma)^{12}C$ \citep{fynbo}
 \item $^{12}C(\alpha,\gamma)^{16}O$ \citep{kunz}.
\end{itemize}

Regions unstable to convection are identified via the Schwartzschild
criterion. We took into account convective overshooting for the core
convective regions and for the surrounding convective shell, assuming
that the turbulent velocity exponentially vanishes outside the
formally convective regions. This is simulated by introducing the
parameter $\xi$, which defines the scale length over which the
velocity decays into the region that is radiatively stable. In this study we
set $\xi$=0.02, in agreement with the calibration given in
\cite{ventura98}.  No overshooting was used for the evolutionary phases
following the core He-burning phase. We note that  reasonable modifications in the overshooting parameter 
 possibly affect the calibration of the absolute ages of star clusters  without a major impact on  the following investigation.

Mass loss [\msun/yr] is modeled according to \cite{blocker}:
\begin{equation}
 \dot{M}=4.83 \times 10^{-22}\eta_RM^{-3.1}L^{3.7}R,
\end{equation}
where $M, L$, and $R$ are denoted in solar units, $\eta_R$ is a free parameter. We used
$\eta_R =0.02$. This value was calibrated specifically for the range of masses
and for the metallicity used here via a comparison between the observed and the
predicted luminosity function of lithium-rich sources and of carbon
stars in the LMC \citep{ventura99,ventura2000}.
However,  variations in this mass-loss rate within the commonly accepted values are not expected to modify our  main results. 

We computed evolutionary tracks for stars with mass from 
$M = 0.4$ $\msun$ to 12 $\msun$,
with an initial chemistry typical of the LMC (Z=0.008) and
$\alpha$-enhanced [$\alpha$/Fe]=0.2, with the reference solar mixture
taken from \cite{gs98}. The evolution of  stars
with both Y=0.25 and Y=0.35 from the pre-MS to the beginning of the
thermal pulses  was calculated (see Fig. \ref{traccerob}).
\subsection{Stellar pulsation models}

For the same chemical compositions as were
adopted in the evolutionary models, we computed a set of nonlinear
convective pulsation models, using the physical and numerical
assumptions discussed in \citet{bms99}, \cite{ v09} and \cite{ m10} for stellar masses
ranging from 3 to 12 $M_{\odot}$. For each stellar mass a
mass-luminosity relation \citep{b00} was adopted\footnote{We note that most of the models for Y=0.25
  are the same as in our previous papers, while the set for Y=0.35 was
  computed specifically for this paper.}, an extensive range in
effective temperature was explored and the modal stability 
was investigated both for the fundamental (FU) and the first-overtone (FO)
instability-strip.  As a result, for each selected chemical
composition, we determined the instability-strip boundaries for both
pulsation modes, as summarized in Table \ref{tbl:1}.  The first four
columns of the table report the adopted chemical composition, mass, 
and luminosity. In the subsequent columns the effective temperature of the predicted 
first-overtone blue edge
(FOBE), fundamental blue edge (FBE), first-overtone red edge (FORE), 
and fundamental red edge (FRE) can be found.

\begin{table*}
\centering
\small
\caption{Physical parameters and location of the fundamental and first-overtone instability-strip boundaries for the adopted pulsation model sets.} %% no full stop at the end of caption
\label{tbl:1}
\begin{tabular}{ccccccccc}
\hline  %% rule at top
Z & Y  & $ M/M_{\odot}$ & $\log{L/L_{\odot}}$ & $T_e(FOBE)$ & $T_e(FBE)$ & $T_e(FORE)$ & $T_e(FRE)$ & source \\
\hline
0.008 & 0.25 & 3.0 & 2.39 &  & 6050 &  & 5850 & This paper \\
%0.008 & 0.25 & 3.0 & 2.64 & 6650  & 6150 & 6050  & 5550 & This paper \\
0.008 & 0.25 & 4.0 & 2.78& 6650 & 5950& 5850 &5625 & Bono et al. 2001 \\
0.008 & 0.25 & 5.0 & 3.07& 6450 & 5950& 5850 &5450 & Bono, Marconi \& Stellingwerf 1999 \\
0.008 & 0.25 & 7.0 & 3.65& 5850 &5850 & 5650 &4950 & Bono, Marconi \& Stellingwerf 1999 \\
0.008 & 0.25 & 7.45 & 3.75 &   & 5850 &   & 4850 & Bono, Marconi \& Stellingwerf 2000 \\
0.008 & 0.25 & 9.0 & 4.00&  &5650 &  &4650 & Bono, Marconi \& Stellingwerf  1999\\
0.008 & 0.25 & 11.0 & 4.40&  &5250 &  &4150 & Bono, Marconi \& Stellingwerf 1999\\
0.008 & 0.35 & 3.0 & 2.59 & 6650 & 6350& 6250 & 5850& This paper \\
%0.008 & 0.35 & 3.0 & 2.84 & 6650 & 6250& 6150 &5650 & This paper \\
0.008 & 0.35 & 4.0 & 3.01 & 6550 & 6150& 6050 & 5650& This paper \\
0.008 & 0.35 & 5.0 & 3.33 & 6350 & 6150& 6050 & 5350& This paper \\
0.008 & 0.35 & 6.0 & 3.60 & 6150 & 6050& 5750 & 5250& This paper \\
0.008 & 0.35 & 7.0 & 3.82 & 5950 & 5950& 5750 & 5050& This paper \\
0.008 & 0.35 & 8.0 & 4.02 &  & 5850&  & 4850& This paper \\
0.008 & 0.35 & 9.0 & 4.19 &  & 5750&  & 4850& This paper \\
0.008 & 0.35 & 10.0 & 4.34 &  & 5750&  & 4640& This paper \\
0.008 & 0.35 & 11.0 & 4.48&  & 5650&  & 4550& This paper \\
0.008 & 0.35 & 12.0 & 4.61&  & 5550&  & 4450& This paper \\
\hline
\end{tabular}
\end{table*}

\begin{figure}
\vspace{-2cm}
\center
\includegraphics[width=1\columnwidth,natwidth=600,natheight=650]{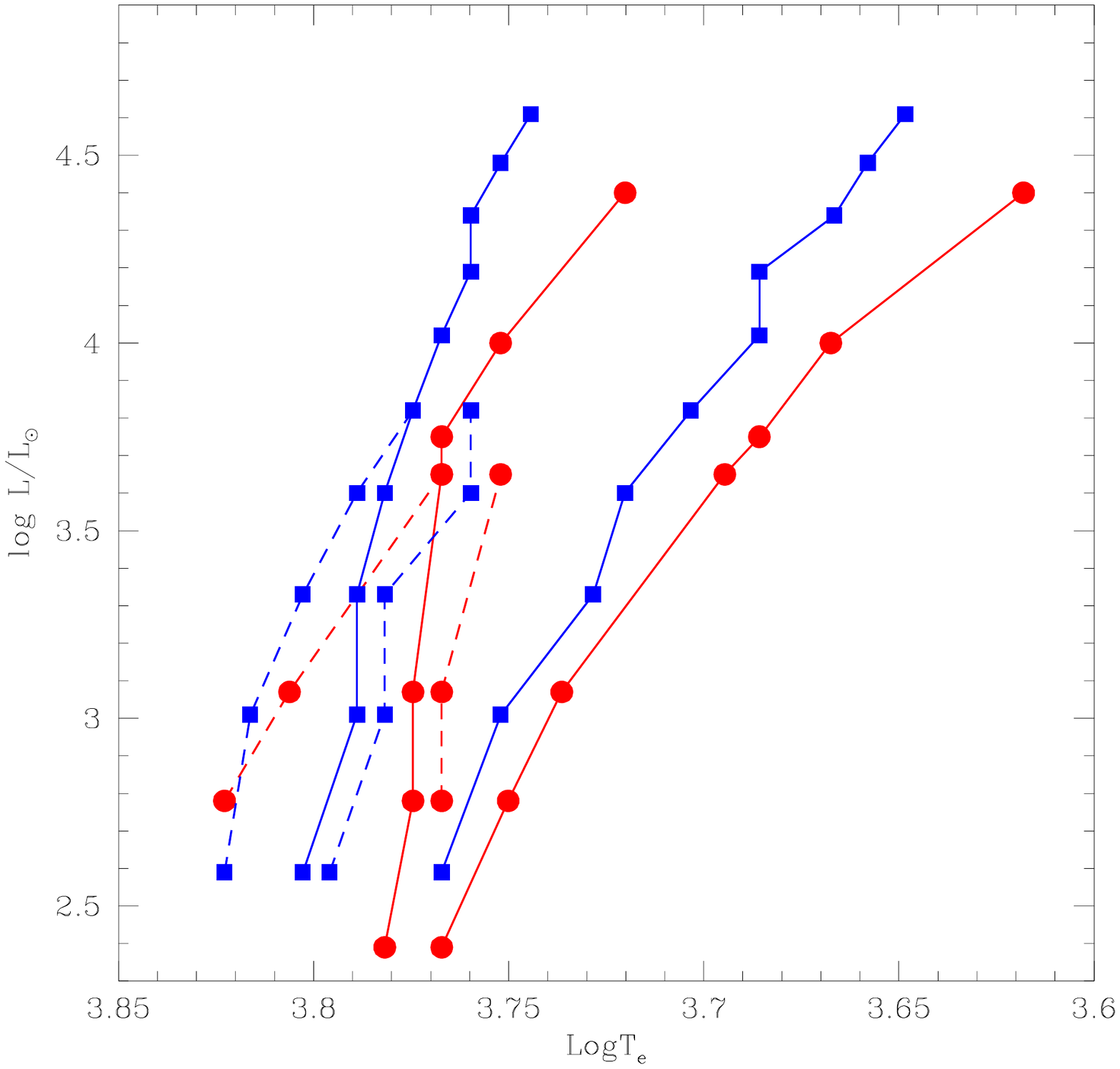}
\vspace{-3cm}
\caption{Comparison between the predicted instability-strip boundaries
for the two assumed helium contents, namely Y=0.25 (red circles) and 0.35 (blue squares). The solid
lines are the fundamental edges, the dashed lines are the first-overtone edges. }
\label{bho}
\end{figure}

A comparison between the theoretical instability-strip for Y=0.25 and
Y=0.35 is shown in Fig. \ref{bho}. As expected on the basis of previous
results \citep{fio02,m05}, the instability-strip of fundamental
pulsators becomes hotter as the helium abundance increases. Here, we also
find that the effect is even more evident for first-overtone pulsators.

To provide a link between the pulsation period and the intrinsic
evolutionary model properties, namely the mass,  luminosity, and 
effective temperature, the pulsation relations were
derived from regression through all the pulsating nonlinear models for
each chemical composition and for both pulsation modes. The
coefficients of these relations are reported in Table \ref{tabl2}.

\begin{figure}[t]
\center
\includegraphics[width=1\columnwidth,natwidth=600,natheight=650]{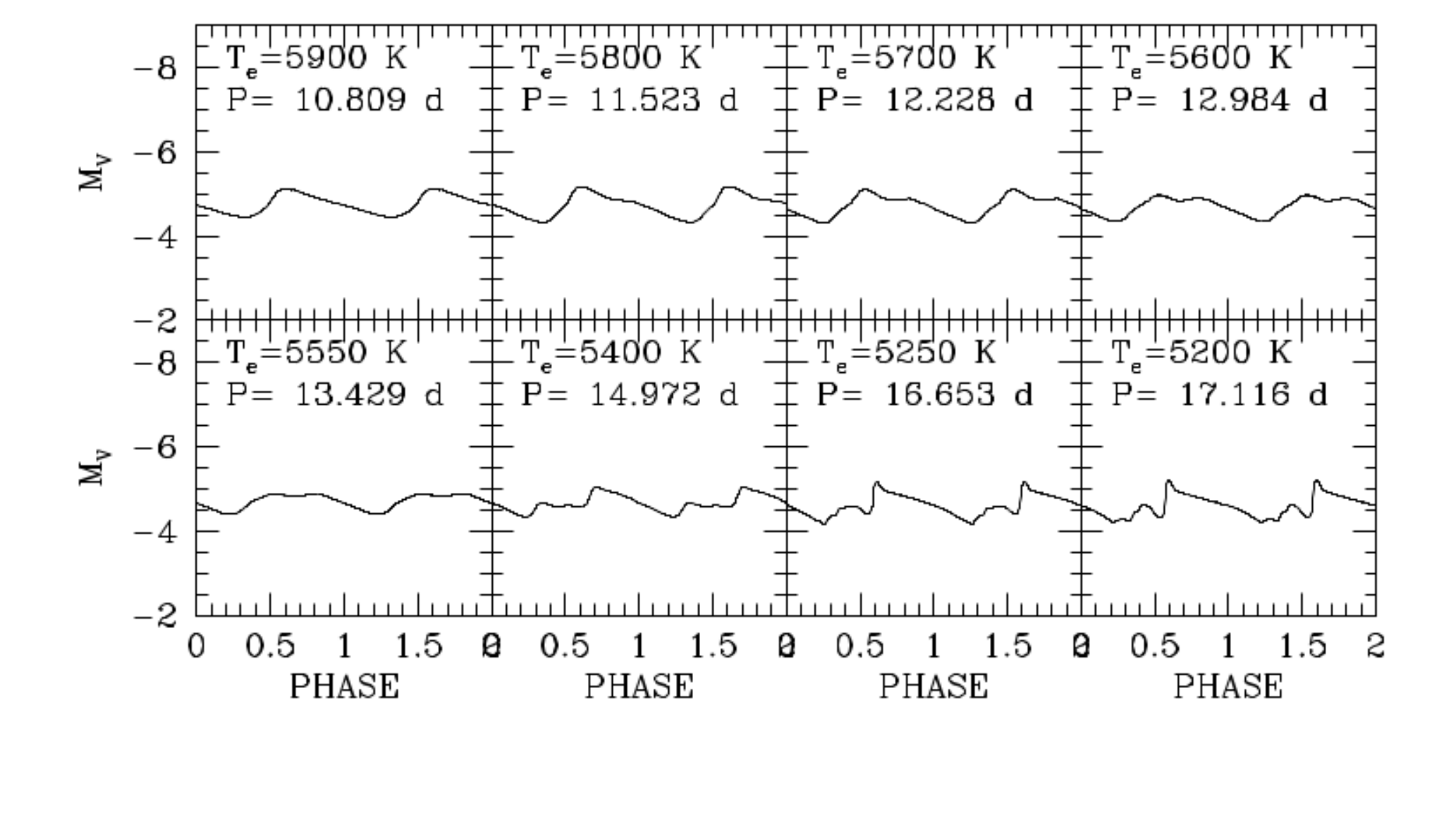}
\caption{Light curves in the absolute V magnitude
%The bolometric light curves
for models with M= $7 M_{\odot}$ and Y=0.35.}
\label{7mass}
\end{figure}

\begin{table*}
\centering
\small
\caption{Coefficients of the adopted pulsation relations: $\log{P} = a
  + b \log{T_e} + c \log{L/L_{\odot}}+ d \log{M/\Msun}$ .}
\label{tabl2}
\begin{tabular}{cccccccc}
\hline  %% rule at top
Z & Y & mode & $a$ & $b$ & $c$ & $d$ & $rms$  \\
\hline
0.008 & 0.25 & FU & 10.556 & -3.279  & 0.931 & -0.795 & 0.008  \\
0.008 & 0.25 & FO & 10.558 &  -3.254&  0.802&  -0.550&  0.003\\
0.008 & 0.35 & FU &  10.741&  -3.344&  0.974&  -0.946& 0.008 \\
0.008 & 0.35 & FO & 10.498 &  -3.245&  0.821&  -0.607&  0.002\\
\hline
\end{tabular}
\end{table*}

The adopted nonlinear pulsation models also allow us to predict the
variations of all the relevant quantities along a pulsation cycle.  A
subset of the obtained light curves in the absolute V magnitude
%bolometric light-curves
is shown in Fig. \ref{7mass} for
models with M=7 $M_{\odot}$ , Z=0.008, and Y=0.35.

The selected stellar mass corresponds to a period range (6 days < P <
16 days) where a secondary maximum (bump) appears both  in the light curve
and in the radial velocity curve. The relationship between the phase of
this bump and the pulsation period is known as the Hertzsprung
progression (HP) \citep{hertz}, and this group of variables has been named
\textit{bump Cepheids}. In previous papers based on the same type of
pulsation models, it has been shown %\citep{bms00,m05}
that an increase in the metal content causes a shift of the HP center
toward shorter periods.  The behavior shown in Fig. \ref{7mass} indicates that
the helium abundance has an opposite effect: the HP center moves to
longer periods because the helium content increases at fixed
metallicity. Indeed, at fixed stellar mass (7 $M_{\odot}$) the center
of the HP occurs at a period of about 13.5 days,
significantly longer than the value obtained at Y=0.25, namely 11.2 d
\citep[see]{bms00}. Empirical data for the LMC
suggest that the HP center corresponds to a period of about 11 days
\citep[see e.g.][]{welch97,beau98},  which agrees better  with the
theoretical value for Y=0.25. However, observations of the Hertzsprung
progression in extragalactic Cepheid samples at the same metallicity
(Z $\simeq$ 0.008) and a similar period range could, in principle, allow
us to distinguish the helium content on the basis of the above
considerations.
This is particularly true in the NIR bands, since \cite{storm}  found a weak dependence on the metallicity in the optical bands. 

\subsection{Stellar population synthesis}

\begin{figure*}
\begin{minipage}{0.47\textwidth}
\resizebox{1.\hsize}{!}{\includegraphics{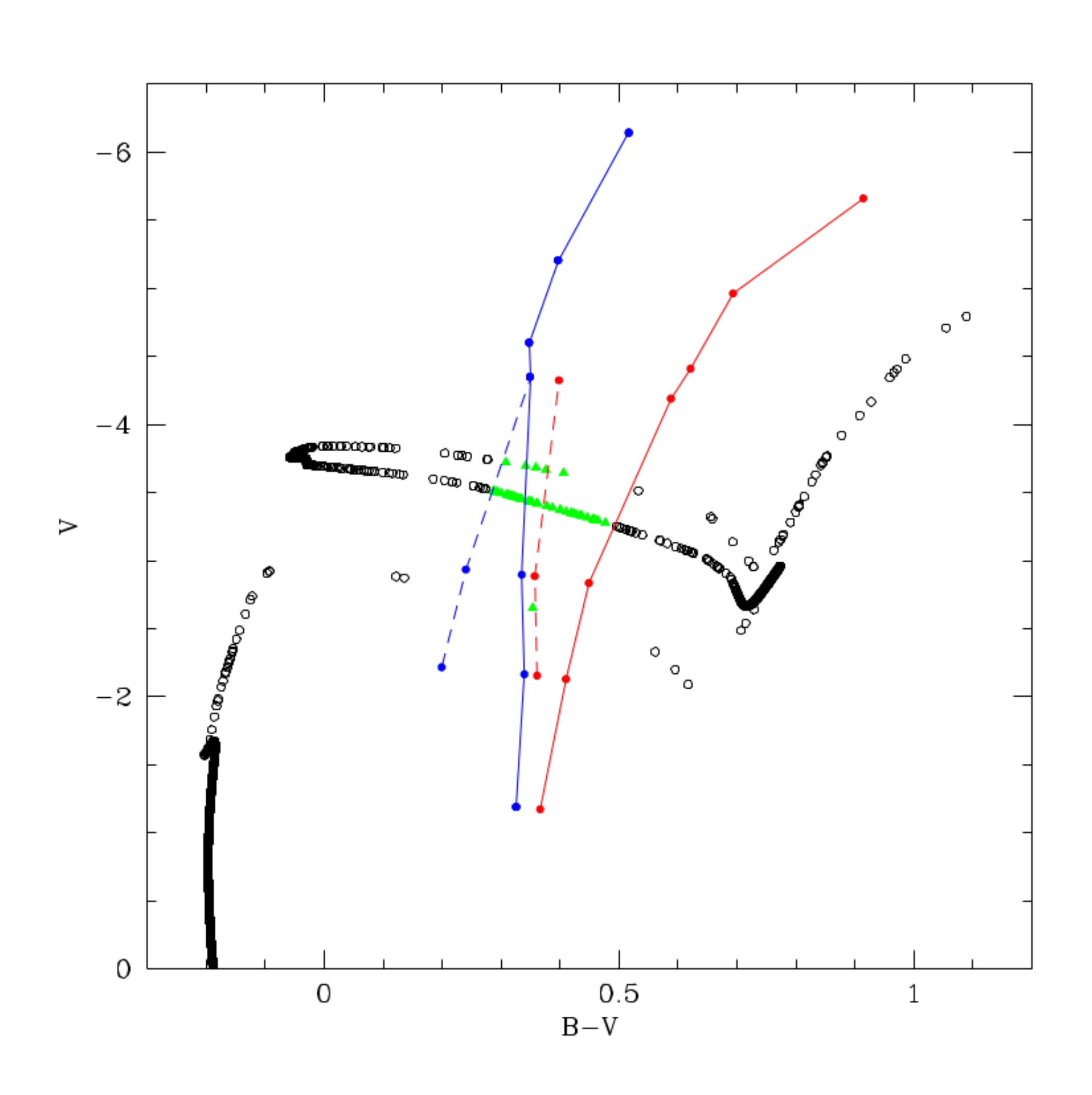}}
\end{minipage}
\begin{minipage}{0.47\textwidth}
\resizebox{1.\hsize}{!}{\includegraphics{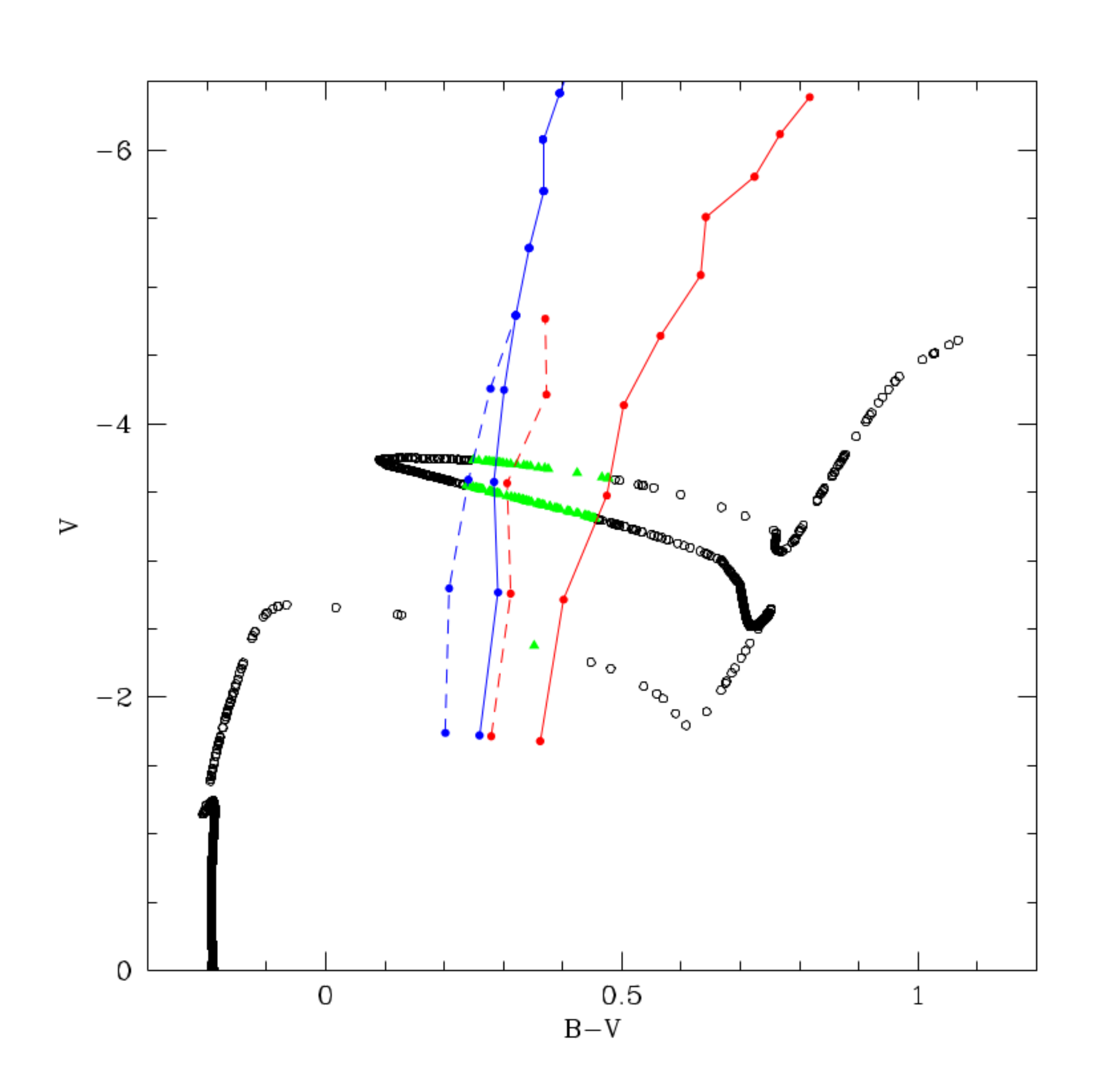}}
\end{minipage}
%\vspace{-0.2cm}
\caption{ Synthetic CMD in the absolute V band for a population of 100 Myr with
  Y=0.25 (left panel)  and Y=0.35 (right panel). The green triangles represent the Cepheids.  The FRE (red line),
   FORE (dashed red line), FBE (blue line) and the FOBE (dashed blue line) are also shown. In these plots the photometric uncertainties are neglected.}
\label{strip}
\end{figure*}

\begin{figure*}[t]
%\vspace{-3.0cm}
\center
\includegraphics[width=1.4\columnwidth,natwidth=600,natheight=650]{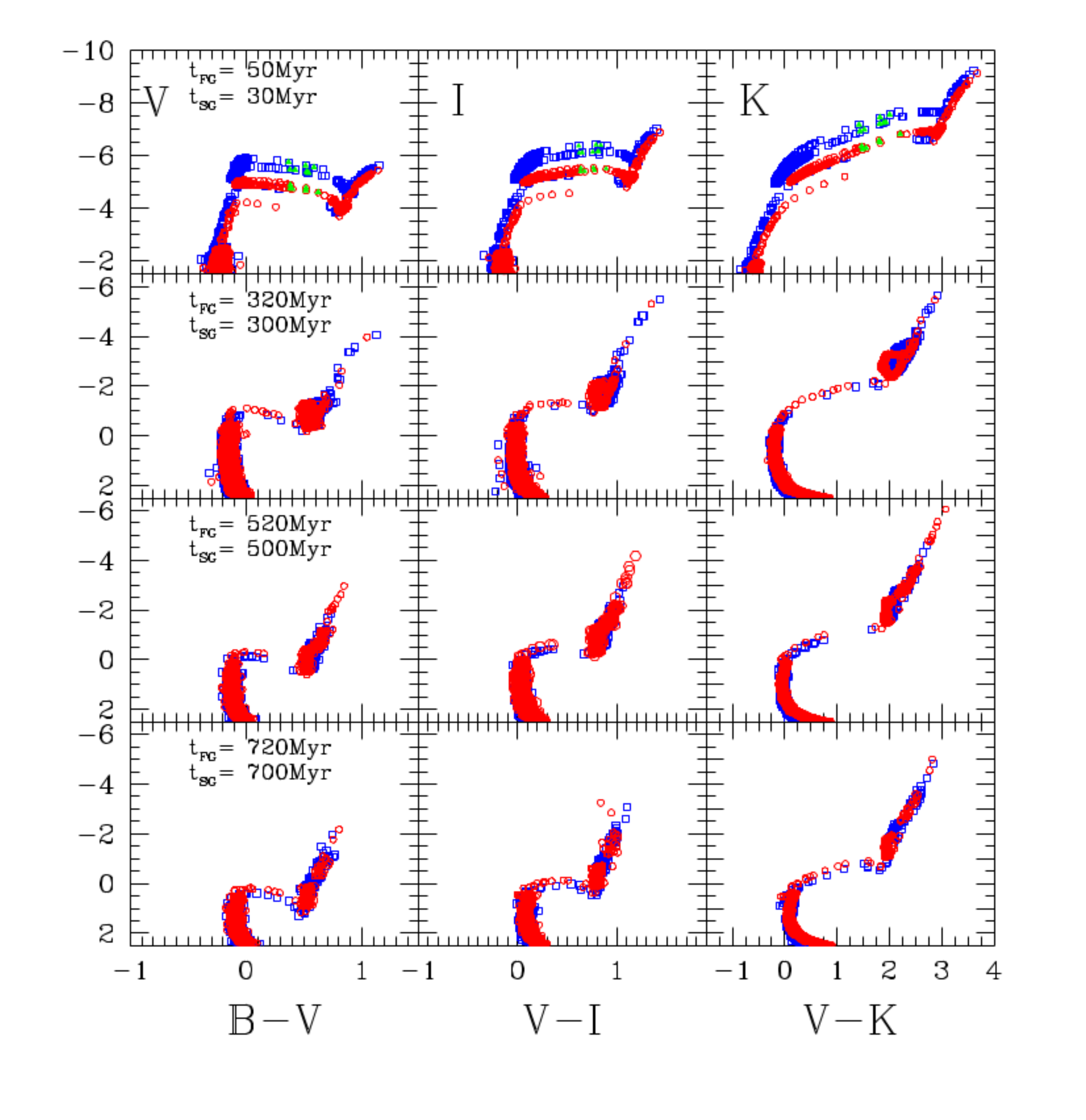}
\caption{Synthetic CMDs in the V, I, and K bands. We hypothesize that the
  GC is composed of two families of stars with Y=0.25 (red circles) 
  and Y=0.35 (blue squares), with an age difference of 20 Myr: the age
  of the first- and second-generation  stars are labeled.  Cepheids
  are marked as green triangles.}
\label{cmddt20}
\end{figure*}

\begin{figure*}[t]
\center
\includegraphics[width=1.4\columnwidth,natwidth=600,natheight=650]{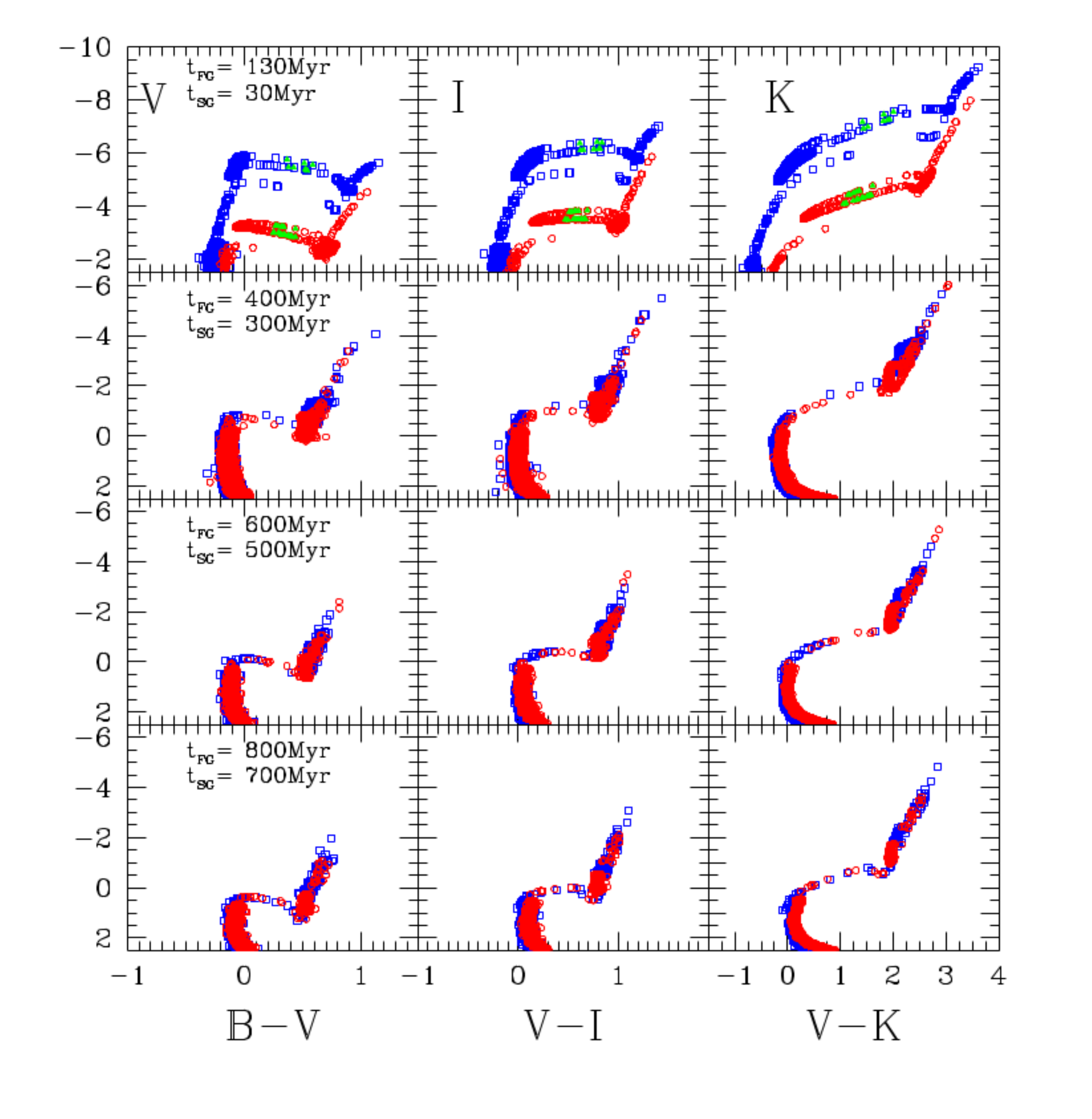}
\caption{As in the previous figure, but for an age difference of 100 Myr between the first- and second-generation  stars. }
\label{cmddt100}
\end{figure*}

Synthetic CMDs and observational properties of the Cepheids were
derived using the most recent version of the stellar population
synthesis code SPoT\footnote{www.oa-teramo.inaf.it/spot}
\citep{2005AJ....130.2625R, 2009ApJ...700.1247R}. The code was used to
fit in detail the observed CMDs of  GGCs and MC
clusters for a wide range of ages and metallicities
\citep{Brocato+00,Raimondo+02,2009ApJ...700.1247R,Bellini+10}. It was used to reproduce the fast and
luminous evolutionary phases and, in particular, the red clump of LMC
star clusters with a high level of accuracy
\citep[e.g.][]{Brocato+03}. For the specific purpose of this study, we
computed synthetic magnitudes and colors (in the UBVRIJHK
bandpasses) by adopting the BaSeL 3.1 \citep{westera,patricelli} stellar atmospheres library. To this aim, the code was implemented to
include the evolutionary tracks calculated with the ATON code (Sect.2.1)
and a specific routine was added to the SPoT code with the aim of
computing the number of predicted Cepheids and, for each variable, the
pulsation mode and the period on the basis of the pulsation models
described in Sect. 2.2.

We adopted a Salpeter initial-mass function (IMF, $\alpha$= 2.35) ranging from 0.4 $\msun$ up to 100 \msun. The total
number of stars in each CMD was chosen to properly populate the
red clump of core He-burning stars. A Monte Carlo method -
weighted according to evolutionary timescale and IMF - was used to
reproduce the stochastic behavior expected in observed
CMDs. Photometric uncertainties were simulated by a specific algorithm
that takes into account the typical increasing spread expected for
photometric CCD measurements of faint stars. We computed synthetic CMDs
and Cepheid properties for two chemical compositions (Z=0.008,
Y=0.25) and (Z=0.008, Y=0.35) in the age range from 20 Myr up to 1
Gyr.
We decided to fix the number of He-burning stars
to be on the order of $\sim 100$ in each model (if not explicitly stated otherwise), so that a well populated core He-burning
phase (i.e. the red/blue clump) was derived.

In Fig. \ref{strip} we show the synthetic [V, B-V] CMD from the MS
to the end of the core He-burning for a population of 100 Myr, with
Y=0.25 (left panel) and Y=0.35 (right panel). Du\-ring the blue loop,
the stars cross the instability-strip, becoming Cepheid variables
(green triangles). At fixed age, the TO point of the He-enriched
population is fainter than the first-generation stars TO point,
since  stars with a higher He abundance evolve more rapidly, while the
blue loop is less extended in color (0.67 vs. 0.8 mag) and spread more
in the absolute V magnitude.
\section{Color-magnitude diagram}

In this section, we focus  on pinpointing some
indicators for multipopulations in young- and intermediate-age stellar systems by using the properties of He-burning stars. It is worth mentioning that these stars are very luminous objects, hence they could represent a powerful way to recognize multigenerations of stars in young clusters provided that they are massive enough to possess a significant number of core He-burning stars.
It is well known that the chemical composition of a star plays a fundamental
role in its evolution. It sets the rate of energy generation and determines the wavelength distribution of the emerging
flux. Stars with a high helium abundance (e.g. Y=0.35), due to the high molecular
weight, evolve in a shorter time and are more luminous and hotter
than stellar models with a lower helium abundance  (e.g. Y=0.25) at fixed mass. For
this reason the MS position and the core He-burning loop of He-rich stellar models are respectively bluer and brighter than
those of classic (i.e. standard He) models with the same
mass (see Fig. \ref{traccerob}). In the following we concentrate  on the expected properties of stars experiencing the core He-burning phase, since the
evolutionary times are long enough to generate a relatively well
populated 'observational' feature in young- and intermediate-age massive stellar systems. 

Figures \ref{cmddt20} and \ref{cmddt100} show a sample of synthetic CMDs (in the V, I, and K bands) for different assumptions on the age of the first- and second-generation stars after fixing their age difference: $\delta t = t_{FG} - t_{SG} = 20$ Myr and 100 Myr, respectively. The ages increase from the top to the bottom of each figure. The helium content is Y=0.25 for the first-generation stars (red circles), while Y=0.35 (blue squares) is assumed for the second-generation stars. The green triangles indicate the Cepheid variables, which we analyze in the next section.

We recall that He-burning stars with mass ranging from $\simeq$ 5 $M_{\odot}$ up to $\simeq$ 10 $M_{\odot}$ experience an extended loop in effective temperature, but they spend a large part of the time needed to exhaust the helium in their convective core at the extreme (cool and hot) edges of the loop. Consequently, we expect to find two clumps at red and blue colors. For the sake of clarify, we define as red a clump located at $(B-V) \geq 0.4$ mag and blue a clump at $(B-V) <0.4$ mag.

We decided to search for photometric indicators for multiple populations using the red and blue clump properties. To this end,  we defined the mean magnitude $<X_{He-b}>$, its dispersion $\sigma_{X_{He-b}}$, and the magnitude spread $\Delta X_{He-b}$ of the red/blue clump as follows:
\begin{equation}
  <X_{He-b}>  = \frac{1}{N_{He-b}} \sum_{i=1}^{N_{He-b}} {X_{He-b}^i}
\end{equation}
\begin{equation}
	 \sigma_{X_{He-b}} = \sqrt {\frac{1}{N_{He-b}}\sum\limits_{i = 1}^{N_{He-b}}  {\left( X_{He-b}^i  - <X_{He-b}> \right)^2 } }
\end{equation}
\begin{equation}
 \Delta X_{He-b} = | X_{He-b}^{max} - X_{He-b}^{min} |,
\end{equation}
where ${X_{He-b}^i}$ is the magnitude of the $i$-$th$ core He-burning star, measured in the given $X$-$band$.
The sum is extended to all the core He-burning stars ($N_{He-b}$). When the blue and red clumps are  present  in the CMD, the sum was computed separately for the stars of  each subset. $X_{He-b}^{max}$ and $X_{He-b}^{min}$ are, respectively, the $X$-$band$ magnitude of the brightest and the faintest edge of the He-burning star clump.
In Tables \ref{amp20} and \ref{amp100} we report these quantities for the computed stellar populations.

In young systems ($t_{FG} \lsim $150 Myr) we can distinguish the two populations both at the MS and at the location of stars burning helium in the core. For $\delta t =20$ Myr (Fig. \ref{cmddt20}), the loops produced by the first- (red circles) and the second-generation (blue squares) stars are clearly separated (in magnitude). Therefore, a double stellar population would be easily discovered from the morphology of the red/blue clump.
The magnitude spread ($\Delta X_{He-b}$) ranges from 0.6 to 1.8 in the V band, and from 0.5 to 2.6 in the K band (Table \ref{amp20}). The color extension of the core He-burning loops does not show sizable differences between the single and double populations because the masses experiencing this evolutionary phase are similar.
When we stretch $\delta t$ to 100 Myr (Fig. \ref{cmddt100}), the magnitude spread increases to a value higher than 2 mag (Table \ref{amp100}). Furthermore, the differences in the morphology of the red/blue loop become even wider than for $\delta t =20$ Myr.

When $t_{FG}$ is $\gsim 300$ Myr, we are unable to distinguish the two subpopulations by simply observing the morphology of the red/blue loop in both cases ($\delta t = 20$ Myr and $\delta t = 100$ Myr). At these ages and for the chemical composition assumed, the blue clump disappears, since the evolution models predict a loop less extended in effective temperature
for the typical mass experiencing the core He-burning phase ($M \lsim 4 M_{\odot}$, see Fig. \ref{traccerob}). However, when the two populations are mixed  the magnitude distribution of stars in the red clump is larger than that produced when the two populations are considered separately. In fact, even if the core He-burning stars of the two subpopulations appear quite superimposed in the CMD, a close  look shows that the magnitude spread of the clump increases  a few tenths of magnitude when populations are mixed, as tabulated in Tables \ref{amp20} and \ref{amp100}. As a general result, we therefore find that the quantity $\Delta X_{He-b}$ is an alternative tool to identify multiple stellar populations in YIMCs. This effect is stronger in the K band than in the bluer bands. Accurate photometric
observations of populous stellar systems are needed to enable one to measure the magnitude spread with 
adequate precision (e.g. $\lsim 0.05$ mag) in the V band.

\begin{figure}
%\vspace{-1.5cm}
\center
\includegraphics[width=1.0\columnwidth,natwidth=600,natheight=650]{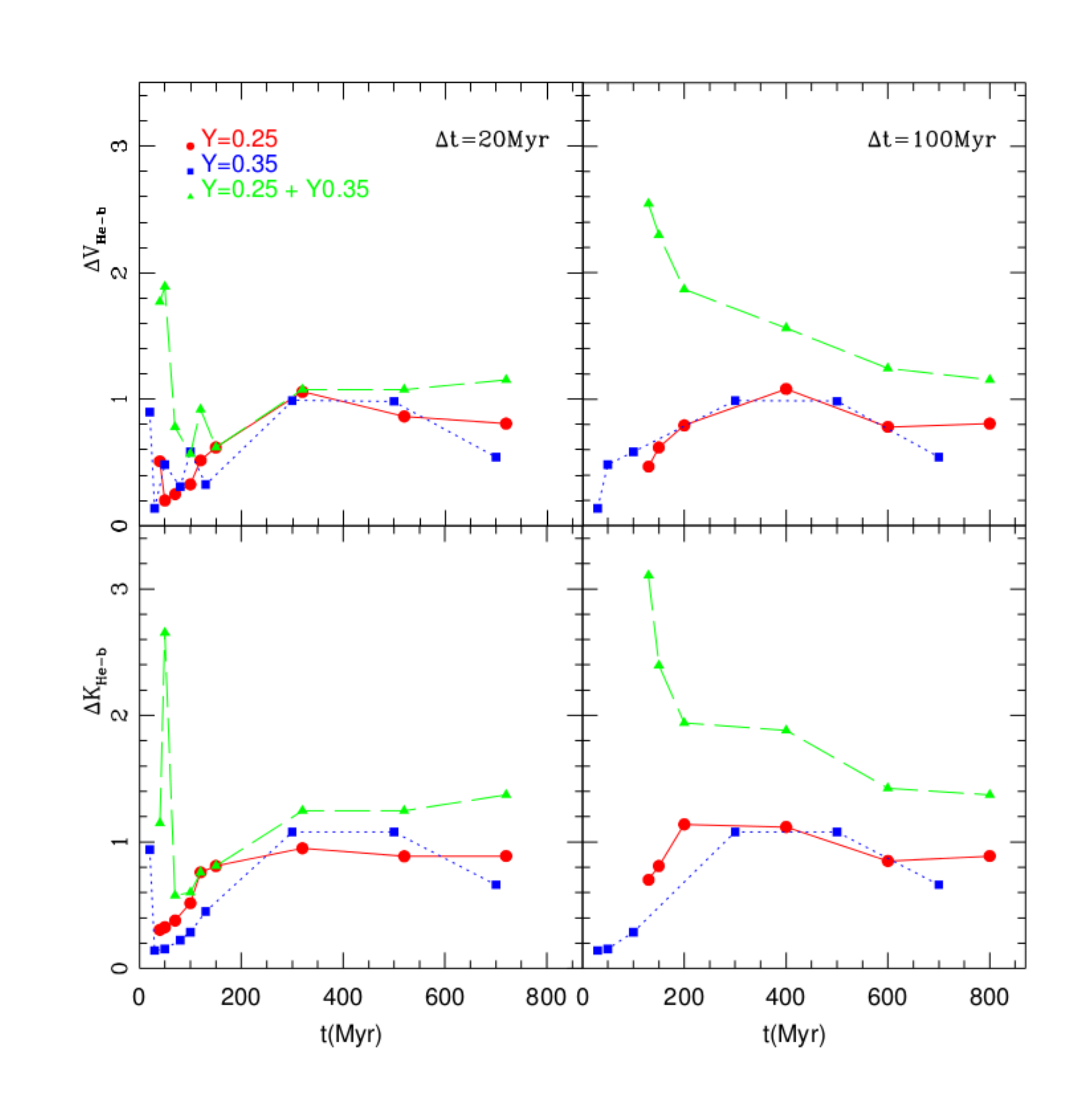}
%\vspace{-3cm}
\caption{
Time evolution of  the $\Delta V_{He-b}$ (top panels) and $\Delta K_{He-b}$
  (bottom panels) bands for a population with Y=0.25 (red full circles), Y=0.35 (blue full squares),
  and a mixed population (green full triangles). The age spread is 20 Myr (left panel) and
  100 Myr (right panel), respectively.}
\label{deltaclump}
\end{figure}

To illustrate this result, in Fig. \ref{deltaclump} we plot the time
evolution of the clump magnitude spread in the V and K
bands. In the figure we omit the I band because  its trend is
similar to that in the V band. Left panels of the figure refer to $\delta t =20$ Myr, while right
panels show $\delta t= 100$ Myr.  The red filled circles and the blue squares describe the magnitude spread of the core He-burning
stars (red clump) for a single population with Y=0.25 and Y=0.35, respectively.  The green triangles
represent the amplitude of the quoted spread when the two populations coexist.

From the figure, it is clear  that the quantity $\Delta X_{He-b}$ is an alternative tool to single out multiple stellar populations in YIMCs.
The magnitude spread changes sizably from single-age to multiple-age (double) populations in the whole range of ages considered here.
In particular, when $\delta t= 100$ Myr (right panels of Fig. \ref{deltaclump}), $\Delta X_{He-b}$ clearly is  a reliable indicator of the possible presence of mixed
populations, at least under the hypotheses explored in the present work.
For young clusters, for example, for $t_{FG}$=130 Myr, $\Delta X_{He-b}$ of the multiple population is
expected to be about four times higher than the value for the single-age popu\-lation in the V and K bands. Increasing the age, this diffe\-rence decreases, because
the mass of stars evolving in the core He-burning phase becomes similar in the two subpopulations. It is interesting to note that this feature is effective in distinguishing   multipopulations in different photometric bands. 

This quantity is less effective when the age difference shrinks to $\delta t = 20$ Myr. The synthetic CMDs
show that the clump of He-burning stars in young clusters ($ t \lsim 150$ Myr) is still a powerful way to evaluate whether there are multiple stellar populations. 
For older ages, the effectiveness of the V band decreases while the K band remains nearly as effective also at ages older than 300 Myr. 

On these bases, we suggest that the magnitude spread $\Delta X_{He-b}$ is a  useful quantity to identify whether there are mixed populations in relatively young- 
and intermediate-age massive stellar systems. We recall  we  assumed age differences between the first- and the second-generation  stars according to the current scenarios proposed to explain multipopulations in star clusters.

We note that before one can compare  this  magnitude spread  with observations,  some refinements are needed. We suggest that $ X_{He-b}^{max}$ and $X_{He-b}^{min}$ should be obtained from the accurate  luminosi\-ty function of  the  red/blue clump stars.
	The faint  edge of the red clump  can be  identified as a   sudden rise  in stellar counts due to the very few stars expected at lower luminosities. On the other hand,    RGB and AGB stars could in principle hamper the defi\-nition of the bright edge. Nevertheless,  the evolutionary time  differences  between the RGB/AGB and the He-burning phases  enable one to predict a sizeable drop ($ X_{He-b}^{max}$) in the luminosi\-ty function.
	For instance, for a 5 $\msun$ star the time ratios are $\tau_{He-b}/\tau_{RGB}$ $\sim$ 15 and $\tau_{He-b}$/$\tau_{AGB}$ $\sim$ 100.
 This allows one to se\-parate the red clump stars from the  RGB/AGB stars. 
The blue clump spread can also be easily   distinguished in the luminosi\-ty distributions. If we analyze the luminosity distribution of the  stars with B-V $<$ 0.4 mag, we can see in the histogram a primary and  a secondary  maximum. The former represents   MS stars, the latter,   at higher luminosities, corresponds to the blue loop stars.  This  secondary feature allows us  to obtain the limits of the magnitude spread. 

We briefly discuss the possibility that a large part of the FG stars are lost  in the very early stage of the cluster life. This clearly affects our statement of having $\sim$ 100 He-burning stars in both the populations. In this case, the magnitude spread remains a good indicator of multiple populations if the number of the FG He-burning stars is on the order of a few tens at least.  In fact,  just a handful of these stars is needed to define the lower edge of  $\Delta X_{He-b}$. Furthermore, this faint side   has  the advantage of not being  affected by the uncertainties due to the presence of RGB and AGB stars.

Before concluding our analysis, we note that  binary systems are not expected to affect  our main conclusions. 
Although observations indicate that the majority of stars are found in binary systems \citep{raste}, close encounters involving binary systems may
disrupt soft (i.e. generally wide) binaries in dense clusters \citep{heggie}. Thus, the actual binary fraction in YIMCs
is  uncertain \citep{zwart,vanbeveren}. Binary stars could, in principle, resemble an increase of the clump magnitude spread. However, their 
contribution to $\Delta X_{He-b}$ would be at most as large as $\sim$ 0.75 mag, but only if the mass difference between the two components is smaller than $\sim$ 0.2 \Msun. 
If so, they can experience the core He-burning phase simultaneously.
If the mass difference is larger than this value, the binary system is most likely composed of a MS star plus a red-clump star.  This type of binaries does not change  $\Delta X_{He-b}$.
Thus, the effect of binaries on the amplitude of the red  clump
will be confined to a handful of stars located at bright
magnitudes (e.g. $\delta V$ $\leq 0.75$ mag). 

%Before concluding our analysis, we note that the ce of binary systems should not affect the main conclusion
%of this paper.  The binary stars could resemble the increase of the amplitude of the red/blue clump, however, the contribution of binaries to %$\Delta X_{He-b}$ would be at most as large as $\sim$ 0.75 mag. On the contrary, in several simulations we find that the increment of $\Delta X_{He-b}$ due to the helium-enhanced
%sub-populations is larger than this value. In addition, the two components of the binary systems should have very similar masses ($\sim$ 0.2 \Msun of difference)
%in order to experience the core He-burning phase simultaneously. If the mass difference is larger the binary system is likely composed by a MS star plus a red-clump star, and it is located %in a region of the CMD ranging from the MS to the red clump, it would not contribute to $\Delta X_{He-b}$.
%Thus, the effect of binaries in the amplitude of the red giant clump
%will be confined to a handful of stars located at slightly brighter
%magnitudes (e.g. $\delta V$ $\leq 0.75$ mag).

\begin{table*}
\center
\small
\caption{Spread of the  (red and blue) clumps in the V, I and K bands for populations with Y=0.25 (FG), Y=0.35 (SG), and the mixed population (multiple). The age difference between the first- and  second-generation  stars is 20 Myr.}
\label{amp20}
\begin{tabular}{ccccccccccc}
\hline
 $t$         & clump & V$_{mean}$ &$\sigma_{V}$ & I$_{mean}$ & $\sigma_{I}$ & K$_{mean}$ & $\sigma_{K}$ & $\Delta V_{He-b}$ & $\Delta I_{He-b}$ & $\Delta K_{He-b}$ \\
 Myr         &       & mag        & mag         & mag        & mag         & mag        & mag          & mag       & mag       & mag \\
\hline
\hline
$t_{FG}$=40  & red  & -4.414 & 0.081 & -5.577 & 0.075 & -7.426  & 0.085 & 0.509 & 0.408 & 0.305 \\
$t_{SG}$=20  &      & -5.590 & 0.246 & -6.546 & 0.061 & -7.952 & 0.341 & 0.897 & 0.288 & 0.939 \\
multiple     &      & -4.661 & 0.498 & -5.781 & 0.401 & -7.537 & 0.276 & 1.769 & 1.270 & 1.148 \\
\hline
$t_{FG}$=40  & blue & -5.299 & 0.044 & -5.401 & 0.067 & -5.462 & 0.116 & 0.223 & 0.304 & 0.469 \\
$t_{SG}$=20  &      & -6.121 & 0.225 & -6.195 & 0.257 & -6.184 & 0.324 & 0.833 & 1.017 & 1.349 \\
multiple        &      & -5.926& 0.401& -6.007& 0.407& -6.013& 0.422& 1.359& 1.507& 1.737 \\
\hline
$t_{FG}$=50  & red  & -3.996 & 0.052 & -5.134 & 0.066 & -6.944 & 0.091 & 0.200 & 0.248 & 0.326 \\
$t_{SG}$=30  &      & -4.578 & 0.022 & -5.761 & 0.024& -7.633& 0.031& 0.136 & 0.133 & 0.142 \\
multiple        &      & -4.301 & 0.334 & -5.463 & 0.365 & -7.306 & 0.410 & 1.891 & 2.203 & 2.653 \\
\hline
$t_{FG}$=50  & blue & -4.926 & 0.046 & -5.082 & 0.107 & -5.225 & 0.240 & 0.227& 0.485 & 1.026\\
$t_{SG}$=30  &      & -5.311 & 0.169 & -5.348 &        0.213 & -5.268 & 0.307 & 0.788 & 1.078 & 1.531\\
multiple        &      & -5.172 & 0.231 & -5.252 &        0.223 & -5.253 & 0.285 & 1.061 & 1.210  & 1.531\\
\hline
$t_{FG}$=70  & red  & -3.421 & 0.063 & -4.517 & 0.078 & -6.253 & 0.108 & 0.249 & 0.286&0.379\\
$t_{SG}$=50  &      & -3.704 & 0.079 & -4.814 & 0.063 & -6.569 & 0.036 & 0.483 & 0.408&0.153\\
multiple        &      & -3.608 & 0.154 & -4.713 & 0.156 & -6.462 & 0.165 & 0.778 & 0.704&0.576\\
\hline
$t_{FG}$=70  & blue & -4.389 & 0.034 & -4.545 & 0.050 & -4.693 & 0.105 & 0.137 &0.242&0.400\\
$t_{SG}$=50  &      & -4.774 & 0.064 & -4.890 & 0.076 & -4.973 & 0.127 & 0.307 &0.435&0.688\\
multiple        &      & -4.678 & 0.176 & -4.804 & 0.165 & -4.904 & 0.172 & 0.666 &0.762&0.947\\
\hline
$t_{FG}$=100 & red  & -2.790 & 0.090 & -3.834 & 0.112 & -5.482 & 0.159 & 0.326 &0.388&0.516\\
$t_{SG}$=80  &      & -2.968 & 0.052 & -4.030 & 0.050 & -5.705 & 0.054 & 0.307 &0.340&0.223\\
multiple        &      & -2.901 & 0.111 & -3.956 & 0.124 & -5.621 & 0.152 & 0.568 &0.594&0.602\\
\hline
$t_{FG}$=100 & blue & -3.761 & 0.041 & -3.942 & 0.040 & -4.128 & 0.052 & 0.148 &0.174&0.284\\
$t_{SG}$=80  &      & -4.125 & 0.040 & -4.360 & 0.038& -4.621 & 0.052 & 0.154 &0.158&0.239\\
multiple        &      & -3.919 & 0.185 & -4.123 & 0.211 & -4.342 & 0.249 & 0.507 &0.580&0.768\\
\hline
$t_{FG}$=120 & red  & -2.476 & 0.119 & -3.494 & 0.146 & -5.089 & 0.204 & 0.516 &0.596&0.760\\
$t_{SG}$=100 &      & -2.623 & 0.115 & -3.647 & 0.098 & -5.250 & 0.071 & 0.585 &0.455&0.286\\
multiple        &      & -2.567 & 0.136 & -3.588 & 0.140 & -5.188 & 0.158 & 0.919 &0.843&0.760\\
\hline
$t_{FG}$=120 & blue & -3.413 & 0.075 & -3.614 & 0.089 & -3.826 & 0.153 & 0.280&0.356&0.592\\
$t_{SG}$=100 &      & -3.710 & 0.077 & -4.073 & 0.050 & -4.513 & 0.092 & 0.287& 0.305&0.387\\
multiple        &      & -3.556 & 0.167 & -3.834 & 0.241 & -4.155 & 0.366 & 0.568&0.742&1.153\\
\hline
$t_{FG}$=150 & red  & -2.098 & 0.157 & -3.080 & 0.188 & -4.603 & 0.262 & 0.618&0.642&0.810\\
$t_{SG}$=130 &      & -2.172 & 0.068 & -3.173 & 0.082 & -4.733 & 0.121 & 0.326&0.323&0.450\\
multiple        &      & -2.138 & 0.124 & -3.130 & 0.149 & -4.672 & 0.210 & 0.618&0.642&0.810\\
\hline
$t_{FG}$=150 & blue     & -3.006 & 0.053 & -3.264 & 0.067 & -3.554 & 0.090 & 0.192&0.243&0.339\\
$t_{SG}$=130 &      & -3.161 & 0.056 & -3.667 & 0.061 & -4.309 & 0.079 & 0.203&0.220&0.258\\
multiple        &      & -3.073 & 0.096 & -3.450 & 0.213 & -3.908 & 0.395 & 0.348&0.660&1.180\\
\hline
$t_{FG}$=320 &  red    & -0.911 & 0.317 & -1.717 & 0.306 & -2.881 & 0.285 & 1.058&1.023&0.95\\
$t_{SG}$=300 &      & -0.868 & 0.292 & -1.735 & 0.303 & -3.015 & 0.323 & 0.988&1.072&1.079\\
multiple        &      & -0.884 & 0.303 & -1.728 & 0.304 & -2.963 & 0.316 & 1.077&1.538&1.247\\
\hline
$t_{FG}$=520 & red     & 0.111  & 0.268 & -0.685 & 0.271 & -1.835 & 0.279 & 0.864&0.888&0.887\\
$t_{SG}$=500 &      & -0.030 & 0.273 & -0.861 & 0.286 & -2.078 & 0.307 & 0.984&1.007&1.079\\
multiple        &      & 0.021  & 0.279 & -0.797 & 0.293 & -1.989 & 0.319 & 1.078&1.156&1.245\\
\hline
$t_{FG}$=720 &  red    & 0.550  & 0.249 & -0.251 & 0.257 & -1.415 & 0.260 & 0.807&0.846&0.889\\
$t_{SG}$=700 &      & 0.027  & 0.133 & -0.821 & 0.151 & -2.068 & 0.174 & 0.541&0.596&0.662\\
multiple        &      & 0.369  & 0.330 & -0.447 & 0.353 & -1.64  & 0.389 & 1.152&1.251&1.374\\
\hline
\hline
\end{tabular}
\end{table*}

\begin{table*}
\center
\small
\caption{Spread of the  (red and blue) clumps in the V, I and K bands for populations with Y=0.25 (FG), Y=0.35 (SG), and the mixed population (multiple).
The age difference between the first- and second-generation stars is 100 Myr.}
\label{amp100}
\begin{tabular}{ccccccccccc}
\hline
 $t$         & clump & V$_{mean}$ &$\sigma_{V}$ & I$_{mean}$ & $\sigma_{I}$ & K$_{mean}$ & $\sigma_{K}$ & $\Delta V_{He-b}$ & $\Delta I_{He-b}$ & $\Delta K_{He-b}$ \\
 Myr         &       & mag        & mag         & mag        & mag         & mag        & mag          & mag       & mag       & mag \\
\hline
\hline
 $t_{FG}$=130& red & -2.355 &0.126 & -3.3591&0.150 &-4.925&0.215&0.467&0.516&0.701\\
$t_{SG}$=30& &-4.578&0.022&-5.761& 0.024&-7.633&0.031&0.136&0.133&0.142\\
multiple& &-3.802&1.062 &-4.923&1.220& -6.689&1.297&2.545&2.716&3.108\\
\hline
 $t_{FG}$=130& blue &-3.278&0.059 &-3.470&0.081&-3.669& 0.125&0.232&0.305&0.494\\
$t_{SG}$=30& & -5.311	&0.169&-5.348&	0.213&-5.268	&0.308&0.788&1.079&1.531\\
multiple & &-4.888&  0.839 &-4.957 &0.787&-4.936 &0.706  & 2.760&2.818&2.964\\
\hline
$t_{FG}$=150& red& -2.098 &0.157& -3.080&0.188&-4.603&0.262&0.618&0.642&0.810\\
$t_{SG}$=50& & -3.704& 0.079&-4.814&0.063&-6.569&0.036&0.483&0.408&0.153\\
multiple& &-3.149&0.772&-4.214&0.834&-5.889&0.948&2.296&2.313&2.395\\
\hline
$t_{FG}$=150& blue&  -3.006 & 0.053 & -3.264 & 0.067 & -3.554 & 0.090 & 0.192&0.243&0.339\\
$t_{SG}$=50& &-4.774& 0.064&-4.900&0.074&-4.973&0.127&0.307&0.4351&0.689\\
multiple& &-4.145&0.846 & -4.322&0.781 &-4.493&0.694&2.160&2.058&1.992\\
\hline
$t_{FG}$=200& red& -1.577&0.240&-2.529&0.282&-3.993&0.368&0.793&0.913&1.139\\
$t_{SG}$=100& & -2.623&0.115 & -3.647&0.098& -5.250&0.071 &0.585 &0.455&0.286\\
multiple&&-2.352&0.485&-3.358&0.517&-4.924 &0.585&1.869&1.888&1.941\\
\hline
$t_{FG}$=200& blue&-2.319&0.128  &-2.812 &0.117&-3.425& 0.134&0.446&0.402&0.595\\
$t_{SG}$=100 & &-3.710 &0.077&-4.073 &0.050&-4.513&0.092 &0.287& 0.305&0.387\\
multiple&&-2.900&0.695&-3.338&0.629&-3.879 &0.549&1.776&1.608&1.515\\
\hline
$t_{FG}$=400& red & -0.401&0.290  &-1.199&0.290& -2.355&0.295  &1.081&1.130&1.118\\
$t_{SG}$=300&& -0.868&0.292  & -1.735 &0.303  &-3.015 &0.323  &0.988&1.027&1.079\\
multiple & &-0.657 &0.373  &-1.493   &0.399 & -2.717&0.452 &1.563&1.712&1.882\\
\hline
$t_{FG}$=600&red & 0.302&0.240 &-0.496&0.239& -1.648 &0.244&0.781&0.794&0.849\\
$t_{SG}$=500& &-0.030&0.273 & -0.861&0.286   & -2.078&0.307 &0.984&1.007&1.079\\
multiple & &0.097&0.306  &-0.721&0.322&-1.914&0.353&1.243&1.305&1.426\\
\hline
$t_{FG}$=800&red & 0.552&0.249& -0.247&0.255&-1.411&0.258&0.807&0.846&0.889\\
$t_{SG}$=700&&0.027&0.133 &-0.821&0.151&-2.068&0.174 &0.541&0.596&0.662\\
multiple&&0.370&0.330 &-0.446&0.353&-1.639&0.389&1.152&1.251&1.374\\
\hline
\hline
\end{tabular}

\end{table*}

\section{Stellar pulsation}

Using  the pulsation period relations as a function of the
intrinsic stellar parameters given in Table \ref{tbl:1}, we computed the period
corresponding to each 'star'
% point
of the synthetic CMDs  
falling within the predicted Cepheid instability
strip  for the two assumed chemical compositions.
Synthetic stars falling between the first-overtone blue edge and the
fundamental blue edge were assumed to pulsate in the first-overtone
mode, and the relative pulsation relation was adopted.
When a synthetic star is within the 'OR region'\footnote{The OR is the region in the instability-strip between FBE and FORE.},
we assumed that the pulsation mode is fundamental if the star, in its evolutionary  path,  crosses the CMD from the red to the blue side. In contrast,
the first-overtone  pulsation was assigned to synthetic stars evolving
from the blue side. Finally, the fundamental-mode is assumed for synthetic stars located between the first-overtone red edge and the fundamental red edge.

\begin{figure*}
\begin{minipage}{0.47\textwidth}
\resizebox{1.\hsize}{!}{\includegraphics{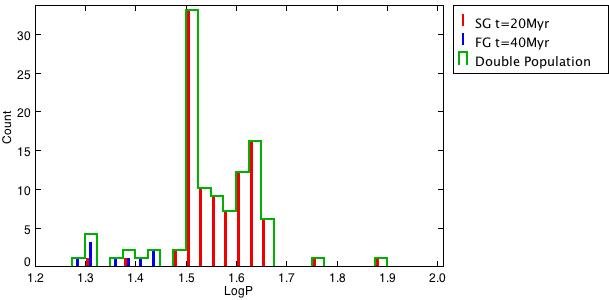}}
\end{minipage}
\begin{minipage}{0.47\textwidth}
\resizebox{1.\hsize}{!}{\includegraphics{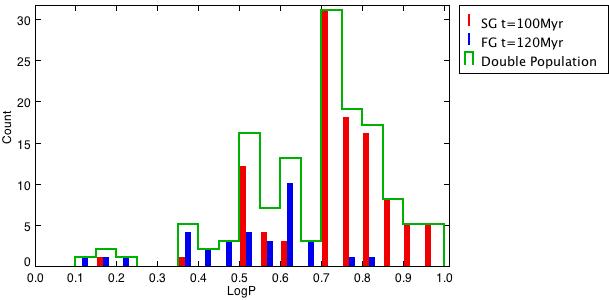}}
\end{minipage}
%\vspace{-1.7cm}
%\vspace{-0.3cm}
\caption{Period distribution of Cepheids for a stellar population containing stars with two different helium abundances (Y(FG)=0.25 and
Y(SG)=0.35), and an age difference of 20 Myr. The age of the first-generation stars is $t_{FG}$ = 40Myr (left panel) and $t_{FG}$=120 Myr (right panel). The green  line  represents the period distribution expected by the double population and defines the adopted bin size (left panel: 0.025; right panel: 0.05). For clarify, the bins of  the red and blue columns are drawn smaller than the adopted ones. }
\label{isto20}
\end{figure*}

When a mixed population -with the first generation of Y=0.25 and t=40 Myr
and the second one  of Y=0.35 and t= 20 Myr- is considered, the resulting
period distribution (left panel of Fig. \ref{isto20}) shows that the
maximum number of pulsators have a period of about 30 days and the 
majority of Cepheids belongs to the second generation and have a
period in the range $\sim$30 $\div$ $\sim$ 45 days.  On the other
hand, if we consider the same age gap, but with the first generation at
120 Myr and the second one at 100 Myr, a mixed Cepheid population is
expected (see the right panel of the same figure) for periods shorter than
about 5 days, whereas a predominance of the second generation is found
between 5 and 10 days.

\begin{figure*}
\begin{minipage}{0.47\textwidth}
\resizebox{1.\hsize}{!}{\includegraphics{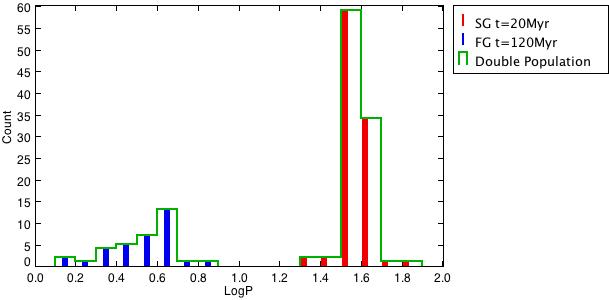}}
\end{minipage}
\begin{minipage}{0.47\textwidth}
\resizebox{1.\hsize}{!}{\includegraphics{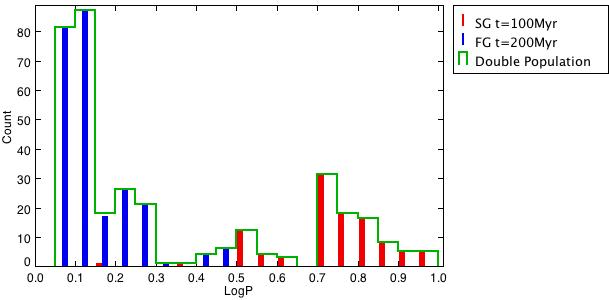}}
\end{minipage}
%\vspace{-1.7cm}
%\vspace{-0.3cm}
\caption{As in Fig. \ref{isto20}, but the age difference between the two subpopulations is 100 Myr. The age of the first-generation stars
is $t_{FG}$ = 120 Myr (left panel, adopted bin : 0.1) and $t_{FG}$=200 Myr (right panel, adopted bin: 0.05). }
\label{isto100}
\end{figure*}

If the second generation is young, as in Fig. \ref{isto20}, but the
first one is older by 100 Myr, we expect the behavior shown in
Fig. \ref{isto100}. We note that by increasing the age gap, the covered
period range becomes wider. In particular, if the Y=0.25 population has an
age of 120 Myr and the Y=0.35 population has 20 Myr, we expect two distinct
period distributions peaked at $\sim$ 5 and $\sim$ 30 days (left panel
of Fig. \ref{isto100}), respectively, with no pulsator in the period
range 8 $\div$ 20 days.  If the Y=0.25 population has an age of 200
Myr and the Y=0.35 population has 100 Myr, we
predict
% expect
a more continuous
distribution with an excess of first-generation Cepheids at the
shortest periods and a predominance of second-generation pulsators for
periods longer than 3 days.

The present exploratory work suggests that an another  tool for distinguishing
 multipopulations (with wide differences in their original helium content) in young massive stellar systems is embedded in
the period distribution of their Cepheids. In fact, our simulations show that the observed period distributions of Cepheids that belong to two different populations
(with age and helium abundances assumed in this paper) tend to considerably increase the range of the measured periods.\\
Thus, a careful analysis of the period distribution of Cepheids in massive stellar systems may give an indication whether there are multiple populations present.
In this way, the detection  of Cepheids with a period longer than expected is another tool for distinguishing multipopulations.

\subsection{Application to the distance scale: the synthetic Wesenheit relations}

It is  well known that the period-luminosity (PL) relation of classical
Cepheids in the optical bands is affected by systematic uncertainties
related to the intrinsic dispersion because of the finite width of the
instability-strip, nonlinearity at the longest periods, reddening, and
chemical composition \citep{bccm99,cmms00}.  To remove the reddening
effect and reduce both the influence of the instability-strip topology
and the metallicity contribution, the Wesenheit relations
\citep{Madore82} are often adopted in the literature.  
Indeed, when two bands are
available, for example, B and V, one can define $WBV=V-R_V(B-V)$,
where $R_V$ is the visual extinction-to-reddening ratio
$A_V/E(B-V)$. The resulting period-Wesenheit (PW) relation is
independent of reddening. Similar PW relations can be defined by
using an extinction law, for instance the one by \citet{Cardelli89}. As the
effect of the extinction is similar to that produced by the finite
width of the instability-strip, the scatter around the PW
relations is smaller than in any observed PL relation \citep[see
e.g.]{CMM00,m05,b10}.  In particular, the PW relation defined using
the V and I bands is also almost independent of metallicity, and hence
represents an ideal tool for estimating the distance of extragalactic
Cepheids \citep{b10}.  In this context, we recall that \cite{storm} found  a  metallicity effect on the coefficients of the PW relations in the optical bands.
\begin{table*}
\centering
\small
\caption{Coefficients of the adopted PW relations for fundamental-mode pulsators: 
$ W_{bands}= \beta\log{P} +\alpha$}
\label{tabl3}
\begin{tabular}{ccccccc}
\hline  %% rule at top
Z & Y & bands & $\beta$ &$\sigma(\beta)$ &$\alpha$ & $\sigma(\alpha)$ \\
\hline
0.008 & 0.25 & BV & -3.39 &   0.01& -2.15  & 0.01\\
0.008 & 0.25 & VI & -3.26 & 0.01& -2.64 & 0.02\\
0.008 & 0.25 & VK &  -3.36 &0.01&  -2.64 & 0.01\\
0.008 & 0.25 & JK & -3.39& 0.01& -2.66 & 0.01\\
0.008 & 0.35 & BV & -3.55 & 0.01& -1.84 &  0.01\\
0.008 & 0.35 & VI & -3.46 & 0.01&-2.33  &0.01 \\
0.008 & 0.35 & VK & -3.52 &0.01&  -2.34 & 0.01\\
0.008 & 0.35 & JK & -3.52& 0.01 & -2.35 & 0.01\\
0.008 & 0.25+0.35 & BV & -3.51 & 0.01& -1.92&0.01\\
0.008 & 0.25+0.35 & VI &-3.40  & 0.01&  -2.42&0.01\\
0.008 & 0.25 + 0.35 & VK & -3.48&0.01 & -2.42 &0.01\\
0.008 & 0.25 +0.35 & JK &-3.50 & 0.01 & -2.43 & 0.02\\
\hline
\end{tabular}
\end{table*}
We can evaluate the synthetic PW relations by simulating stellar populations with different original helium content.
For these computations we assumed for each given age a total cluster mass of about $5\times 10^5$ $\Msun$.
The ages were chosen to populate the Cepheid period range (0.4 $\lsim$ LogP $\lsim$ 2) typically adopted in the empirical evaluation of
the Wesenheit relation \citep{udalski,b10}.
The coefficients of the resulting relations for fundamental-mode
pulsators are reported in Table \ref{tabl3} for each adopted chemical composition.
First of all, we note that our WVI
relation for the case with Y=0.25 agrees very well  with the one obtained from
observations of LMC Cepheids \citep{udalski}. Moreover,  our coefficients of the WJK, WVK, and WVJ relations   reproduce  the recent results of \cite{inno} very well, although we  note some discrepancy with their WVI relation.
When the helium content increases from Y=0.25 to Y=0.35  the   Wesenheit
relations become steeper.
\begin{figure*}
\center
\includegraphics[width=.90\columnwidth,natwidth=600,natheight=650]{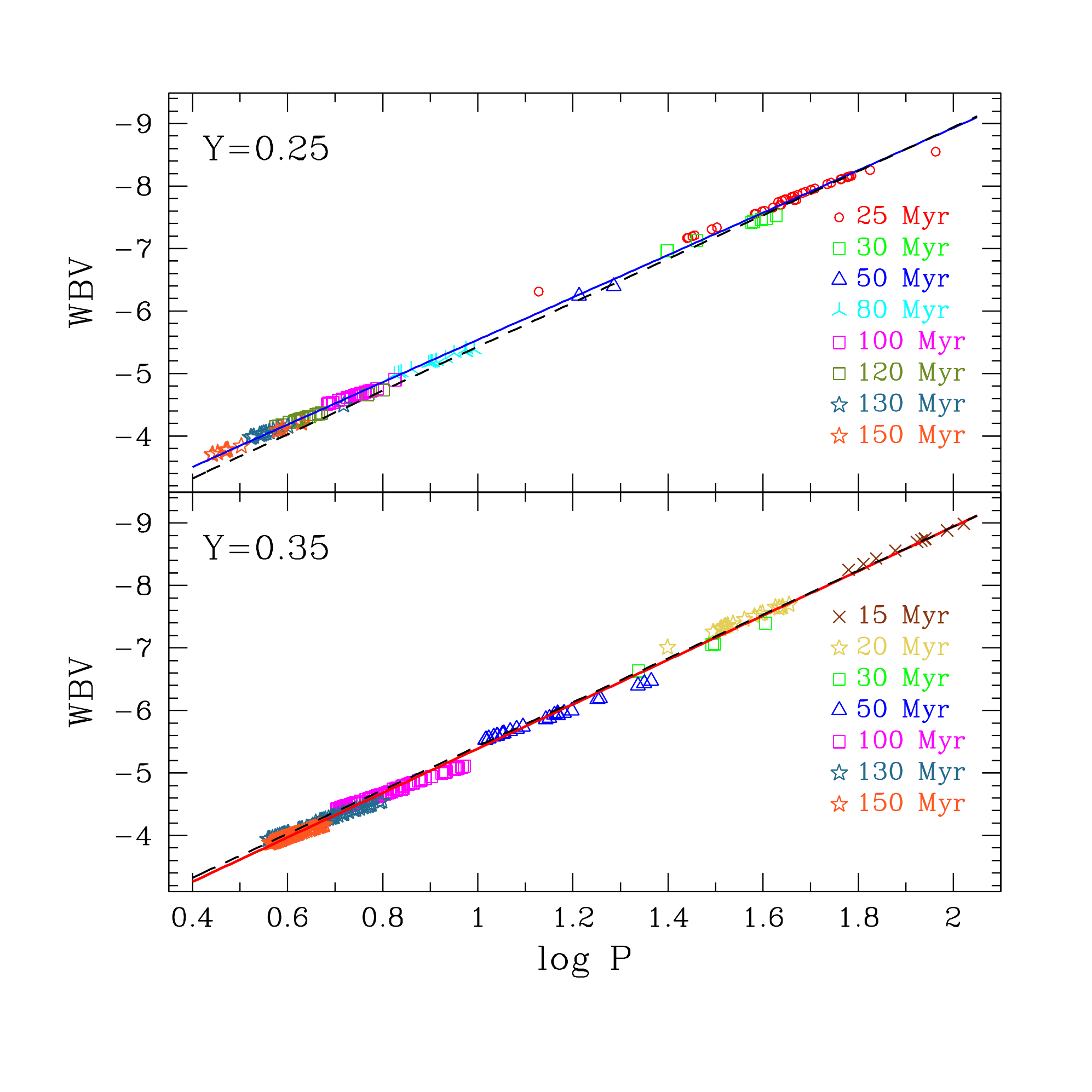}
\includegraphics[width=.90\columnwidth,natwidth=600,natheight=650]{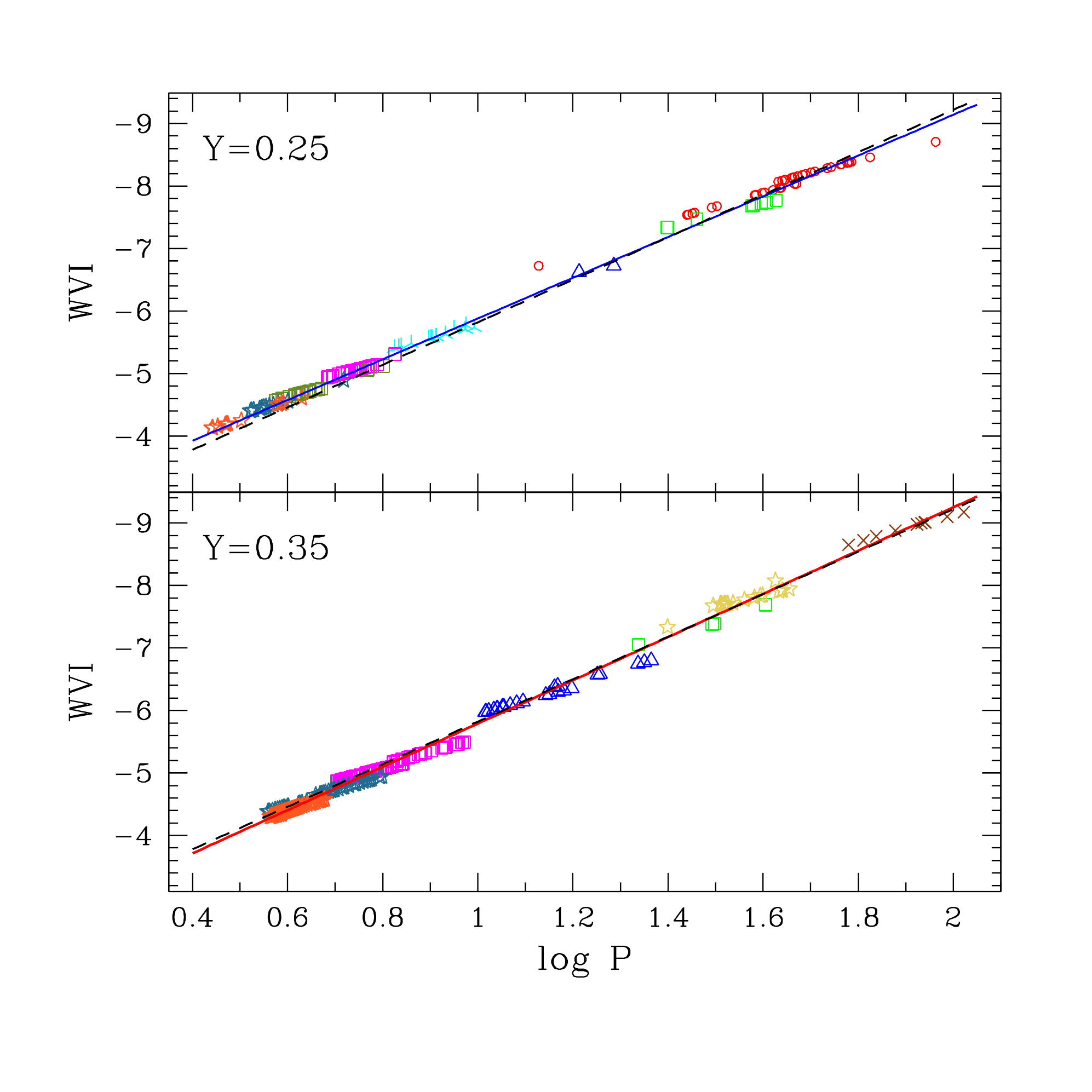}
\includegraphics[width=.90\columnwidth,natwidth=600,natheight=650]{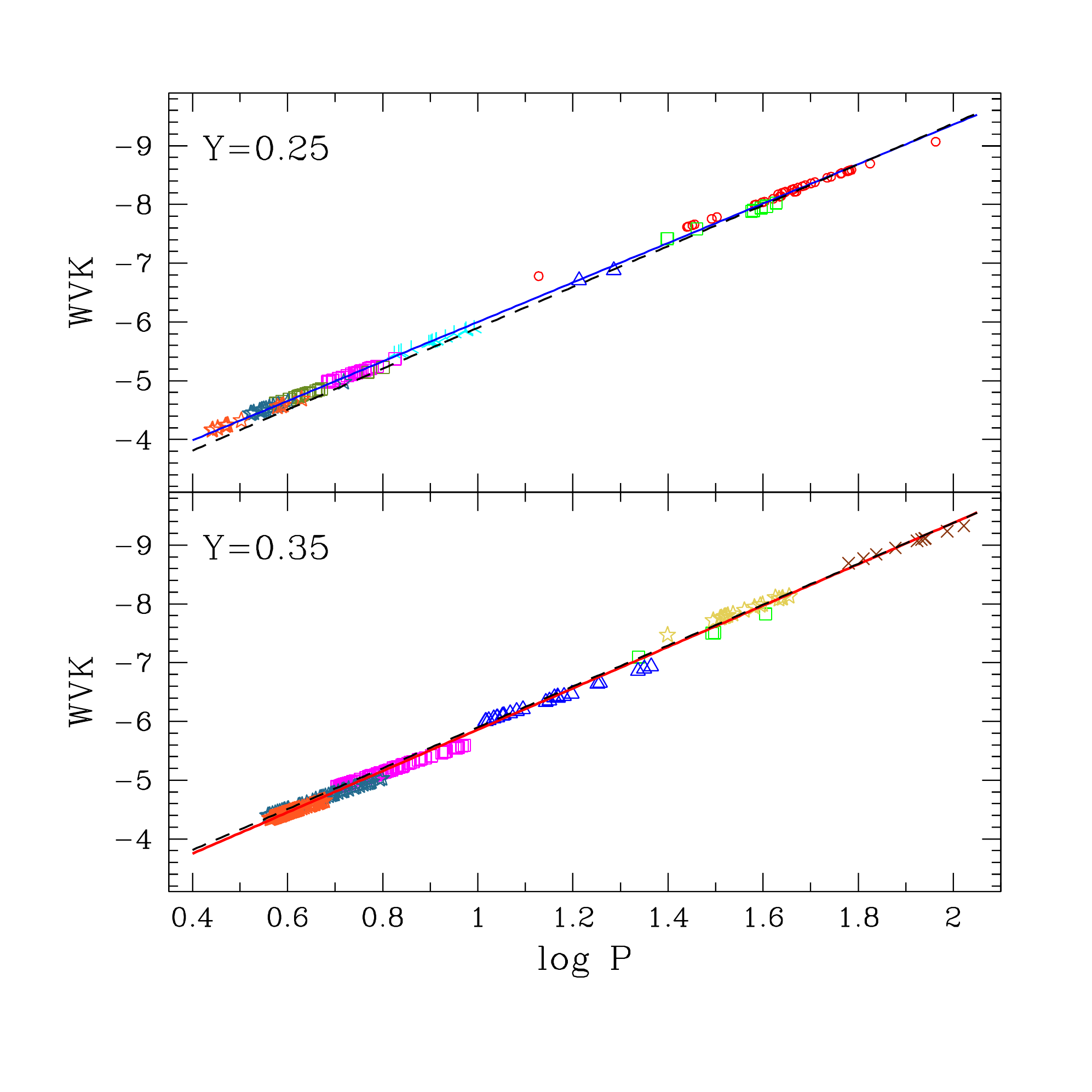}
\includegraphics[width=.90\columnwidth,natwidth=600,natheight=650]{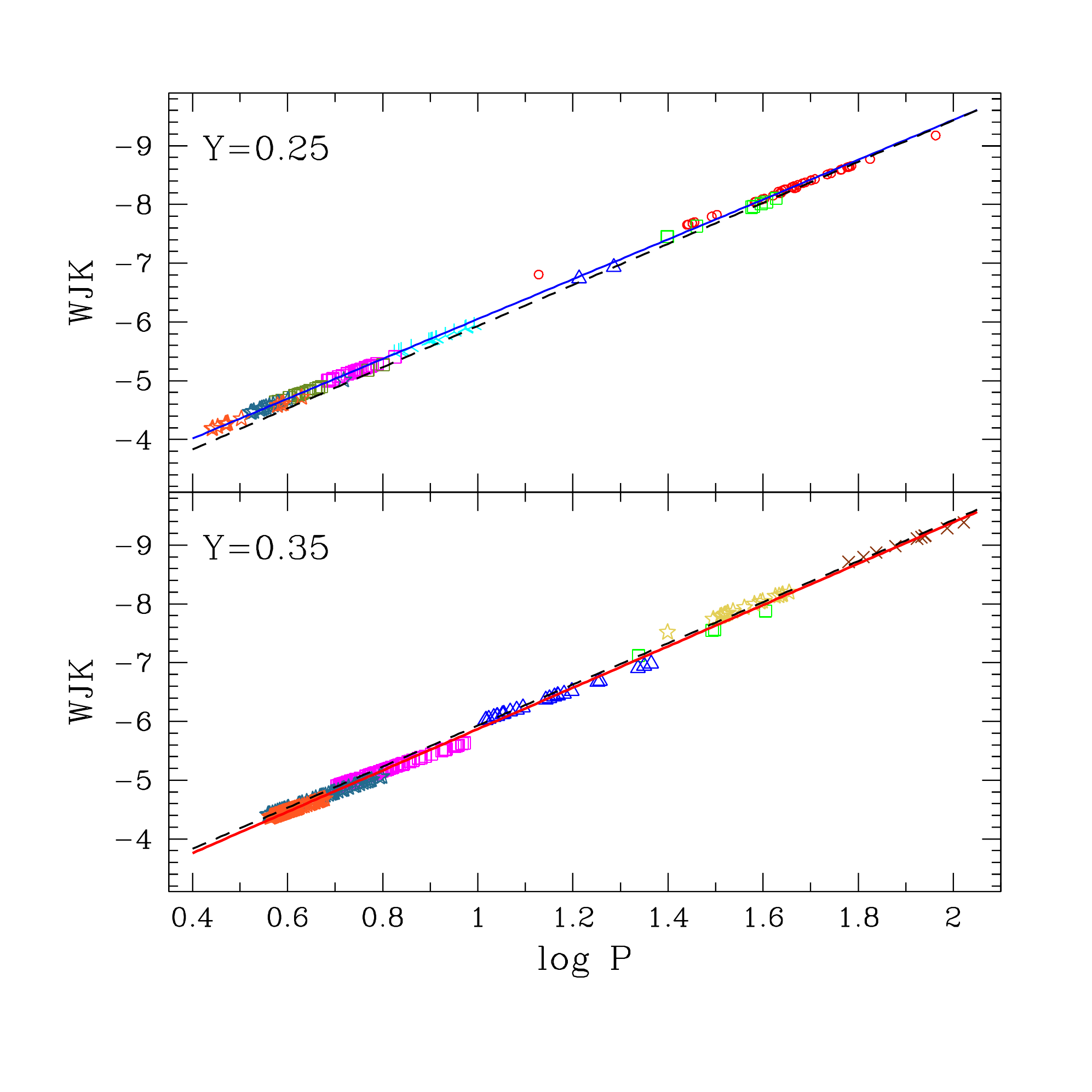}
%\vspace{-2cm}
\caption{PW relations in different filter combinations [V,(B-V)] (upper left panel),  [I,(V-I)] (upper right panel), [K,(V-K)] (lower left panel)
and [K,(J-K)] (lower right panel) for the two single-He abundance populations, as labeled. Colors refer to populations with different ages as
reported in the upper left panel.
 In each upper (lower) panel, the solid blue (red) line is the PW relation derived for Cepheids with Y=0.25 (Y=0.35),
 while the dashed black line indicates the PW relation derived by combining the stellar populations with different helium abundances.}
\label{wk}
\end{figure*}
Figure \ref{wk} shows the synthetic fundamental-mode  Wesenheit
relations for the various filter combinations, namely [V, (B-V)], [V, (V-I)], [V, (V-K)], and [K,(J-K)],  for the two assumptions on the helium
content.
In each panel the dashed line depicts the relation assuming a combined
stellar population with two adopted helium abundances (last four lines of
Table 5).
We note that the combined relations are closer to those for Y=0.35
than to those at Y=0.25. This occurrence is due to the different
period distributions of Cepheids at different helium content,
with the pulsators at Y=0.35 dominating the short-period range.
We note that depending on the period range, the effect can be as large
as 0.2-0.3 mag.
This implies that adopting the LMC-calibrated Wesenheit relation  to derive
the distances of Cepheid samples
belonging to populations
affected by He enrichment mechanisms
may be inadequate.

\section{Conclusions}

We explored the possibility of detecting multiple
populations  in star clusters by observing bright, intermediate-age
massive stars that experience the core He-burning phase. These stars typically gather together, forming a red/blue
clump in the CMD of very populous stellar systems with ages from few Myr up to
$\sim 1$ Gyr. Our work was motivated by the simple consideration that because GGCs
show multiple populations, signs of multiple gene\-rations of stars can probably also be found  in young
stellar systems, under the assumption that the formation process of star
clusters is universal. This view  seems supported by recent observations,
because in a few MC clusters some broadening of the main sequence has been detected
\citep[e.g.][]{milone,milone2013}.

As  a starting step, we focused  on stellar populations with different original helium content, as suggested in recent works on GGCs \citep[e.g.][]{hb,piotto,monelli}
and by theoretical scenarios \citep[]{ventura2001,decressin} proposed to explain the origin of multiple populations.

To this aim, an updated set of stellar evolutionary tracks were
computed with masses  from 0.4 M$_{\odot}$ up to 12 M$_{\odot}$.
In particular, we computed models to
reproduce the synthetic CMDs of stellar populations with a fixed metallicity
similar to stars observed in the LMC (i.e. Z = 0.008) and two different
original helium contents (Y=0.25 and Y=0.35).

With the same physical inputs, a new set of He-enriched stellar pulsation models was
calculated as well. All these ingredients were included in the stellar population synthesis code
SPoT to produce synthetic CMDs for single-helium and mixed-helium abundance populations. We investigated the
hypothesis that two stellar populations differ in age as $\delta t
=20$ and $100$ Myr, the younger population with an helium abundance of Y=0.35, while the older population has Y=0.25.
The new pulsation models allowed us to derive the borders of the instability-strip
and to obtain quantitative information on the pulsation features of variables belonging to
single-age and multiple-age populations.

One of the novelties of this work is that we tackled the problem by using
a homogeneous approach in handling stellar evolutionary tracks and
stellar pulsation models. Both sets of models  were used and were linked
 in the stellar population synthesis code to guarantee
 full consistency.

Under the hypotheses presented above, the main results of the paper may
be summarized as follows:

\begin{itemize}

\item We suggest that the magnitude spread $\Delta X_{He-b}$
($X$ is one of the VIK photometric bands) of
  the red clump is a valuable tool to single out multiple stellar
  populations in massive clusters with age within the quoted
  range.  This amplitude
  assumes different values if we observe a single-age stellar
  population or more than one.   Discovering multiple populations in
  young systems is very important to determine the
  appropriate scenario to explain the formation of multiple
  populations in GCs. 

\item For  double populations, the  
  $\Delta X_{He-b}$  values are expected to be systematically higher than for of a single population.

\item The Cepheid period distribution
  in double populations is broader than that
  expected for a single stellar
  population.

\item The synthetic period-luminosity and Wesenheit
  relations obtained in this work furthermore support the possibility of some bias effects in deriving distances because of the potential presence of double
  stellar populations (with a  large difference in their helium contents) in the sample of Cepheids used to measure their distances.

\item Owing to the age interval considered, massive stars are
  evolving out of the main sequence, thus they are luminous and easy
  to detect at large distances. The drawback is that one needs to
  observe rich star assemblies  to obtain a significant number of stars in this
  bright but fast evolutionary phase. According to our simulations,
  the mass of such  star assemblies should not be lower than $\simeq
  5 \times 10^{5}$ $M_{\odot}$.
  
\end{itemize}

It is worthwhile to clarify that the  work intends to focus  on observational features produced by differences 
in the original helium content and does not intend to exclude the possibility that other quantities (e.g. metallicity) may mimic analogue observable effects. 
Nevertheless, we show that   modifications of the observable $\Delta X_{He-b}$ and  the   Cepheid period distributions,  is expected  if a massive stellar system is composed of two populations with a large difference in their He abundance.
% with respect to what measurable in case of single stellar population.
 
 We establish that double stellar populations
could be identified in very populous YIMCs by using our indicators,
providing that the spatial resolution and the distance of the targets allow
accurate photometric measurements of individual stars. This is expected to be possible thanks to the new
generation of observational facilities (i.e. E-ELT, JWST), since they will push the resolution of young
stars up to the Fornax cluster of galaxies and beyond.
Finally,  we confirm that the extragalactic distance scale may be affected by bias due to unexpected presence of Cepheids whose original He content is very different from the one assumed in the generally adopted $PW$ relationships  \citep[e.g][]{fio02,m05}.

\begin{acknowledgements}
 It is a pleasure to thank the referee for her/his useful suggestions and comments. 
  This work received partial financial support by INAF$-$PRIN$/$2010
  (PI G. Clementini) and INAF$-$PRIN$/$2011 (PI M. Marconi).
\end{acknowledgements}

%\bibliography{manuscriptbib}

             \end{document}